\documentclass[10pt,a4paper]{article}
\usepackage[top=1.25in, bottom=1.5in, left=0.8 in, right=0.8 in]{geometry}

\usepackage{amsmath, amsthm, amsfonts,  amssymb, latexsym}
\usepackage[colorlinks, citecolor=Red, linkcolor=Blue, urlcolor=blue]{hyperref}
\usepackage{caption}
\usepackage[dvipsnames]{xcolor}
\usepackage{tikz}
\usepackage{tensor}
\usepackage{subcaption}
\usepackage{pgfplots}
\usetikzlibrary{arrows,arrows.meta, decorations.pathmorphing, decorations.markings, decorations.pathreplacing,intersections, pgfplots.fillbetween,patterns,shadings}
\usepackage{enumerate}
\usepackage{cancel}
\usepackage{relsize}

\makeatletter
\DeclareRobustCommand{\ibar}{\mathord{%
  \text{$\m@th\mkern-2mu\raisebox{-0.7ex}[0pt][0pt]{$\mathchar'26$}\mkern-7mu i$}%
}}
\makeatother

\newcommand{\ket}[1]{\lvert #1 \rangle}

\newcommand{\del}{\partial}

\newcommand{\beq}{\begin{equation}}
\newcommand{\eeq}{\end{equation}}

\DeclareMathOperator{\sech}{sech}
\DeclareMathOperator{\csch}{csch}

\pgfplotsset{compat=1.18}
\usepackage[colorlinks, citecolor=Red, linkcolor=Blue, urlcolor=blue]{hyperref}

\usepackage{soul}
\usepackage{xcolor}    

\definecolor{mint}{RGB}{189, 252, 201}
\definecolor{mintblue}{RGB}{163, 255, 225}
\definecolor{lightred}{RGB}{255, 220, 220}

\sethlcolor{lightred}  
\begin{document}

\title{Quantum effects in near-extremal charged black hole spacetimes}
\author{Maria Alberti\textsuperscript{1,2,3}\thanks{mariaalberti@campus.technion.ac.il} \ and Jochen Zahn\textsuperscript{1}\thanks{jochen.zahn@itp.uni-leipzig.de} \\ 
\textsuperscript{1}\textit{Institut für Theoretische Physik, Universität Leipzig,}  \\ \textit{Brüderstr. 16, 04103 Leipzig, Germany}\\
\textsuperscript{2}\textit{Max Planck Institute for Mathematics in Sciences (MiS), }  \\ \textit{Inselstraße 22, 04103 Leipzig, Germany}\\
\textsuperscript{3}\textit{Department of Physics, Technion, Haifa 32000, Israel} 
}
\date{April 1, 2025}
\maketitle

\begin{abstract}
 We compute the semiclassical current and stress-energy fluxes both at the event and Cauchy horizon of a near-extremal Reissner-Nordström black hole. We consider a minimally-coupled, massless, weakly charged scalar field in the Unruh state, describing an evaporating black hole. The near-extremal domain allows for an analytical treatment of the scattering problem of the Boulware modes both in the interior and exterior regions. We present this and explicit analytical expressions for $\langle j_v\rangle $, $\langle T_{vv}\rangle$ at the horizons, as well as estimates for $\langle j_u\rangle$ and $\langle T_{uu}\rangle $. We cross-check the analytical results numerically by bringing the radial Klein-Gordon equation into the form of the confluent Heun equation. Inserting these expectation values as sources to the Einstein-Maxwell equations, we find that at least in the near-extremal regime of small field charge, quantum effects drive the black hole interior away from extremality. Our work generalizes the known results for the real scalar field \cite{Zilberman:2021vgz} and is in agreement with recent work on charged fields in expanding Reissner-Nordström deSitter universes \cite{Klein:2021ctt,Klein:2021les}. 
\end{abstract}

\section{Introduction}
Quantum field theory in curved spacetimes (QFTCS) describes the interaction of quantum matter with classical gravitational (and potentially electromagnetic) fields. The coupling of both is governed by the semiclassical Einstein(-Maxwell) equations
\begin{align}\label{semicl EN}
    G_{\nu\rho} & =8\pi \big(\langle T_{\nu\rho}\rangle_\Psi +T_{\nu\rho}^\text{EM}\big), &
    \nabla^\mu F_{\mu \nu} & =-4\pi \langle j_\nu\rangle_\Psi,
\end{align}
where the quantum expectation values of the renormalized stress tensor $T_{\nu \rho}$ and current $j_\nu$ act as additional sources. The field is in a physically reasonable state of choice $\ket{\Psi}$. A non-vanishing expectation value of the quantum observables will influence the classical fields and this will, in turn, influence again the quantum observables. This phenomenon is called backreaction. One does not, in practice, attempt to directly solve \eqref{semicl EN}. In the usual QFTCS approach, one first sets a solution to the classical Einstein(-Maxwell) equations to act as the fixed background. On this background one constructs the quantum theory and evaluates the quantum observables. The effects of these on the geometry are only taken into account a posteriori; the newly influenced geometry is fixed again according to these and the process should be repeated iteratively. 

We will call (certain components of) the expectation value of the stress tensor (or current) energy (or charge) fluxes. The calculation of such fluxes in black hole (BH) spacetimes is relevant in two regimes: In the exterior region, the fluxes of energy and charge due to Hawking radiation (and/or superradiance, see below) drive the evaporation of the BH. This is a process typically occurring on very long timescales, so that as a first approximation, backreaction can be ignored. On the other hand, as conjectured already in the 1970s \cite{Birrell:1978th}, and proven in recent years \cite{Zilberman:2019buh,Hollands:2019whz,Hollands:2020qpe,Klein:2021ctt,Zilberman:2022aum,Klein:2024sdd,McMaken:2024fvq}, energy and charge fluxes generically diverge at the Cauchy horizon inside rotating and/or charged BHs when expressed in coordinates that extend smoothly across the Cauchy horizon. These divergences are typically of higher degree than those due to classical fields, and in particular strong enough \cite{Hollands:2019whz} to rescue strong cosmic censorship in situations in which it would be violated in the context of classical fields \cite{Cardoso:2017soq,Dias:2018ufh}. The occurrence of divergences in the fluxes at the Cauchy horizon also makes it clear that near a Cauchy horizon backreaction effects need to be taken into account. This opens the exciting possibility to study backreaction effects in a regime where they are a major driver of the dynamics and even occur in a situation where the curvature of the background geometry can be arbitrarily small. For this reason, the study of quantum effects in BH interiors has been a major activity in QFTCS in recent years, see for example \cite{Zilberman:2022iij,McMaken:2023tft,Klein:2023urp,McMaken:2024tpc,Zilberman:2024jns} and the references given above.

With the notable exception of \cite{Zilberman:2021vgz}, previous studies of quantum effects near the Cauchy horizon relied on numerical evaluation of scattering coefficients on the BH exterior and interior. Obtaining analytical results is desirable at least for two reasons: i) Analytical results are expected to be beneficial in a more detailed study of backreaction effects. ii) Previous results \cite{Klein:2021ctt} showed that quantum effects can ``upcharge'' a BH interior (instead of the discharge naively expected). One would like to prove that near extremality quantum effects always discharge the BH interior (as indicated by numerical results \cite{Klein:2021ctt}), so that it is not driven above criticality. In \cite{Zilberman:2021vgz}, analytical approximations for the scattering coefficients for the real scalar field on near-extremal Reissner-Nordstr\"om (RN) spacetime were found and corresponding near-extremal approximations for the renormalized stress-energy tensor (SET) at the Cauchy horizon were obtained for the Unruh \cite{Unruh:1976db} and the Hartle-Hawking state \cite{Hartle:1976tp} (the first describing a black hole evaporating by Hawking radiation, the second one a black hole in thermal equilibrium with the quantum field).

Uncharged scalar fields on non-rotating BH spacetimes (as studied in \cite{Zilberman:2021vgz}) are rather special, as they do not exhibit superradiance, which provides an evaporation mechanism \cite{Starobinsky:1973aij,Unruh:1974bw,Bekenstein:1973mi,Gibbons:1975kk} which can be considered as distinct from Hawking radiation (which is thermal). Near extremality, where the Hawking temperature is very small, the evaporation by superradiance is thus expected to dominate over Hawking evaporation, and thus to also contribute the dominant contribution to the energy and further fluxes (charge, for example). It is thus highly desirable to generalize the results of \cite{Zilberman:2021vgz} to a setting including superradiance. We do this here, by considering a charged scalar field on RN. This leads to considerable technical complications compared to \cite{Zilberman:2021vgz}. We obtain analytical approximations for the scattering coefficients in the near-extremal limit under the supplementary condition of small field charge $q$, i.e., $q Q \ll 1$ with $Q$ the BH charge. Using these, we obtain analytical approximations for the charge and energy fluxes at the event and the Cauchy horizon in the Unruh state. Our results for the energy fluxes reduce to those of \cite{Zilberman:2021vgz} in the limit $q \to 0$ of vanishing field charge.

The result for the fluxes at the event horizon suffice, by exploiting the symmetry of the state and conservation of current and stress energy tensor, to determine the charge and energy fluxes at infinity \cite{Balakumar:2022yvx}, and thus the evaporation rates for charge and mass of the BH. In the extremal limit, our results give a universal ratio $4/3$ between these (at leading order in the field charge). Hence, the evaporation drives the BH exterior away from extremality, as expected. 

We also discuss backreaction effects in the interior by studying (in the weak backreaction approximation \cite{Zilberman:2019buh,Klein:2021ctt}) the evolution of the local charge (defined by Gau{\ss} law) and the quasi-local Poisson-Israel mass. We find that their decay rates are also governed by the decay rates at infinity, and thus in particular have the same $4/3$ ratio in the extremal limit. In this sense, also the BH interior is driven away from extremality by quantum effects. 

Our explicit analytical expressions for the fluxes are only valid in the near-extremal approximation and in the regime of small field charge. However, the qualitative features of the analysis are valid for generic field charges, as in that approximation 1) superradiance is still dominant and 2) our results in the interior region are still valid (they only rely on near-extremality). In particular, the scattering in the interior region governs the relationship between the Cauchy and event horizon fluxes, the latter determined exclusively by the scattering outside of the black hole (and dominated by superradiant effects in the near-extremal limit). Therefore, although our analytical analysis is restricted to small field charges, our expectation –supported by numerical results– is that it captures the qualitative features of generic near-extremal scenarios.

The paper is structured as follows: in sec. \ref{setup section} we review the Reissner-Nordström geometry as well as the canonical quantization in the Unruh state of a charged field in this background. In sec. \ref{section: observables} we compile the relevant results available in the literature regarding the explicit expressions of the observables at the horizons as mode-sums in terms of the scattering coefficients in the BH exterior and interior. The number of angular momentum $\ell$ modes for which the scattering problem must be solved, as well as the relevant frequency regime, are not immediately clear. However, by restricting the analysis to the near-extremal regime and assuming a small field charge, we find that the problem simplifies and can be treated analytically. Specifically, we find that under these conditions, the mode-sum is dominated by the $\ell=0$ mode. Moreover, it is not necessary to solve the scattering problem in the exterior region for arbitrary frequencies $\omega$; instead, it suffices to consider the (potentially slightly extended) superradiant regime. The pivotal work of this paper is developed in sec. \ref{section: scattering coeffs}: we present an analytical analysis of the radial Klein-Gordon equation in both the interior and exterior region of the black hole in the near-extremal approximation. 

Building upon this, we compute the semiclassical current and energy fluxes at the event and Cauchy horizon. These results are cross-checked numerically by bringing the radial Klein-Gordon equation into the form of the confluent Heun equation \cite{ronveaux1995heun}. In sec. \ref{section: semiclassical fluxes}, the analytical approximations of the scattering coefficients are used to determine analytical results (in the near-extremal approximation) for the charge and energy fluxes at the event and the Cauchy horizon. The physical consequences of these results regarding physics near the Cauchy horizon are discussed in sec. \ref{section: discussion} in a similar fashion to \cite{Klein:2021ctt,Zilberman:2019buh} in the context of weak-backreaction. We conclude with a summary and an outlook.

Throughout, we work in units $\hbar =c=G=4\pi\epsilon_0=1$.

\section{Setup}\label{setup section}
\begin{figure}
    \centering
    \begin{subfigure}[b]{0.42\textwidth}
        \includegraphics[width=0.75\textwidth]{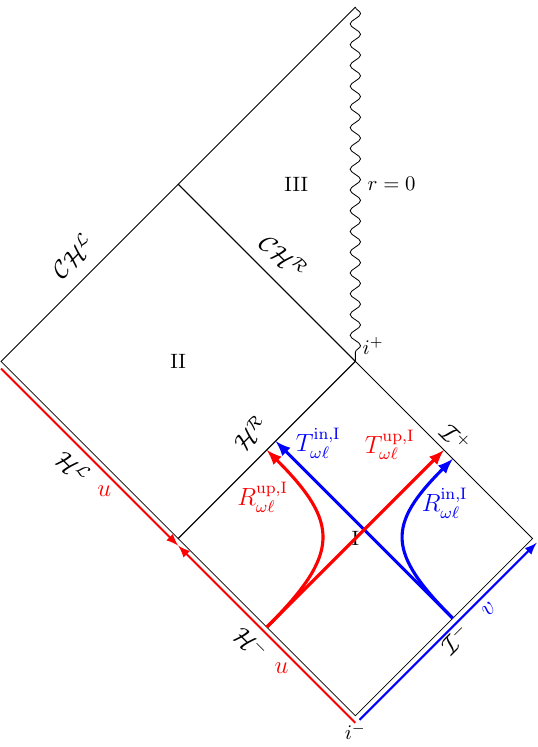}
        \caption{Scattering of the in (blue) and up (red) Boulware modes in region I.}
    \end{subfigure}\hspace{0.5cm}
    \begin{subfigure}[b]{0.42\textwidth}
        \includegraphics[width=0.75\textwidth]{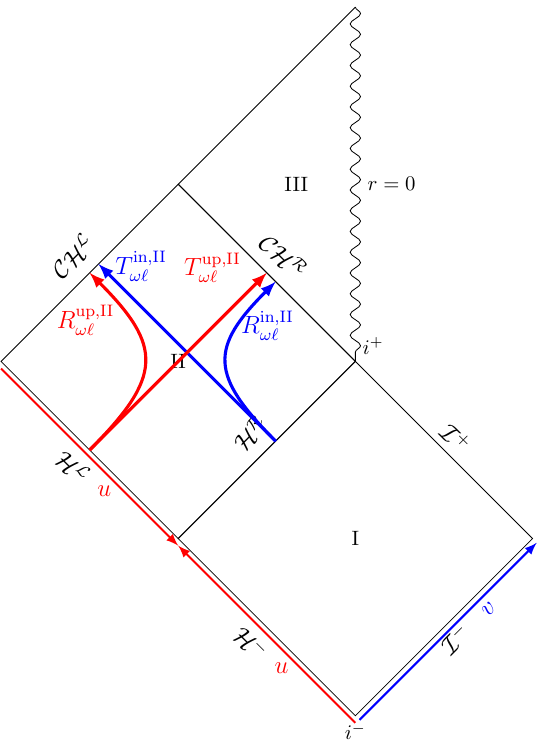}
        \caption{Scattering of the in (blue) and up (red) Boulware modes in region II.}
    \end{subfigure}
    \caption{}\label{fig:scatteringBoulware}
\end{figure}
We consider a minimally coupled complex scalar field propagating in a charged black hole spacetime. The solution to the Einstein-Maxwell equations which describes this setup is, up to a gauge transformation, the Reissner-Nordström solution
\begin{subequations}\label{RN}
\begin{align}
    &g=-f(r)dt^2+f(r)^{-1}dr^2+r^2d\Omega^2, \hspace{1cm} f(r)=1-\frac{2M}{r}+\frac{Q^2}{r^2}\label{metric tr form}\\
    &\hspace{4cm} A=-\frac{Q}{r}dt,
    \end{align}
\end{subequations}
where $M$ and $Q$ refer to the mass and electric charge of the black hole, respectively, $A$ is the background electromagnetic potential and $d\Omega^2=d\theta^2+\sin^2\theta d\phi^2$ denotes the area element of the unit 2-sphere. The charge parameter satisfies $\vert Q\vert <M$ for subextremal BHs and $\vert Q\vert =M$ in the extremal case. The function $f(r)$ has two positive roots $r=r_\pm$, corresponding to the location of the event/outer ($r=r_+$) and Cauchy/inner ($r=r_-$) horizon, which coincide in the extremal limit. The hypersurfaces $\{ r=r_\pm \}$ are Killing horizons with respect to $\partial_t$ and associated surface gravity $\kappa_\pm=\frac{1}{2}\vert f'(r)\vert_{r=r_\pm}$. These coincide in the extremal case. The Penrose diagram for subextremal RN is depicted in Fig. \ref{fig:scatteringBoulware}. The blocks I, II and III are defined as
\begin{equation}
   \text{I}=\mathbb{R}_t\times(r_+,\infty)\times S_{\theta,\phi}^2, \hspace{0.5cm}  \text{II}=\mathbb{R}_t\times(r_-,r_+)\times S_{\theta,\phi}^2,\hspace{0.5cm}\text{III}=\mathbb{R}_t\times(0,r_-)\times S_{\theta,\phi}^2.
\end{equation}
In the following, we refer to region I as the exterior region, to region II as the interior or black hole region and to region III the black hole interior. Regions I and II are separated by the right event horizon $\mathcal{H}^\mathcal{R}$ whereas the other two are separated by a right Cauchy horizon $\mathcal{CH}^\mathcal{R}$. We introduce the tortoise coordinate, defined via $dr_*=f(r)^{-1}dr$ and the double-null coordinates $u=t-r_*$, $v=t+r_*$. In order to extend the metric smoothly across the horizons, one introduces Kruskal coordinates
\begin{equation}
    U\equiv\pm \frac{1}{\kappa_+}e^{-\kappa_+ u} \hspace{1.5cm} V\equiv-\frac{1}{\kappa_-}e^{-\kappa_- v},
\end{equation}
which are regular at the horizons. The $U$ coordinate allows us to extend the metric across the event horizon, from $r>r_+$ ($-$ sign) to $r<r_+$ ($+$ sign). 

We will restrict our analysis to the near-extremal regime. This is quantified by the parameter
\begin{equation}\label{near extremality parameter}
    \Delta=\sqrt{1-Q^2/M^2}=\frac{r_+-r_-}{2M}\ll 1,
\end{equation}
which vanishes in the extremal limit. In terms of this parameter, the location of the horizons is exactly $r_\pm=M\pm M\Delta$ and the surface gravities are given by 
\begin{equation}\label{surface gravity}
    \kappa_\pm=\frac{M\Delta}{r_\pm^2}=\frac{\Delta}{M}+\mathcal{O}\Big(\frac{\Delta^2}{M}\Big),
\end{equation}
in particular they are indistinguishable to leading order and vanish in the extremal limit. We study the charged scalar field subject to the Klein-Gordon equation
\begin{equation}\label{klein-gordon}
     D_\nu D^\nu \Phi =0 \hspace{1cm} D_\nu=\nabla_\nu-iq A_\nu,
\end{equation}
where $q$ is the charge of the field and the covariant derivative is taken with respect to the metric \eqref{RN}. It is customary to use the separation ansatz
\begin{equation}\label{mode ansatz}
    \Phi_{\ell m}(t,r,\theta, \phi)=(4\pi)^{-1/2}r^{-1}Y_{\ell m}(\theta, \phi)H_\ell (r,t)
\end{equation}
for the Klein-Gordon scalar field modes in terms of the spherical harmonics $Y_{\ell m}(\theta, \phi)$. If one additionally assumes 
\begin{equation}\label{t separation}
    H_\ell (r, t)=e^{-i\omega t} h_{\omega \ell} (r),
\end{equation}
the Klein-Gordon equation \eqref{klein-gordon} reduces to an ordinary differential equation on the $r$-variable
\begin{subequations}\label{radial boulware equation}
\begin{align}
    &\partial_{r_*}^2 h_{\omega \ell}^{\Lambda}(r)=\Big\{V
_\ell(r)-\Big( \omega-\frac{qQ}{r}\Big)^2 \Big\}h_{\omega \ell}^{\Lambda}(r)\hspace{0.5cm} \Lambda\in \{ \text{I}, \text{II}\}, \label{radial KG}\\
    &\hspace{1.4cm} V_\ell(r)=f(r)\Big(\frac{\ell(\ell+1)}{r^2}+\frac{\partial_r f(r)}{r} \Big).\label{potential radial KG}
\end{align}
\end{subequations}
 As $t$ (or $r_*$) does not extend beyond $\mathcal{H}^\mathcal{R}$, the solutions $h_{\omega \ell}^{\Lambda}$ only exist in regions I and II separately. The potential $V_\ell$ vanishes exponentially at the horizons when expressed in terms of $r_*$, i.e. $V_\ell\propto e^{\pm 2\kappa_\pm r_*}$. At radial infinity it decays like $\mathcal{O}\big(\frac{1}{r^2}\big)$, whereas the term coupling to the electromagnetic field decays only like $\mathcal{O}\big(\frac{1}{r}\big)$ (or $\frac{1}{r_*}$). In contrast to the real scalar field case, this prevents the radial solution from becoming free at radial infinity.  

The charged scalar field allows for gauge transformations 
$
    A_\nu \to A_\nu+\partial_\nu \chi(x),\ \Phi\to e^{iq\chi(x)}\Phi
$
for arbitrary gauge functions $\chi(x)$. For our purposes it will be most useful to achieve that the electromagnetic potential vanishes at one of the horizons. For this, only gauge functions of the form $ \chi_\pm(x)=\frac{Q}{r_\pm}t$ are necessary. Following the notation in \cite{Klein:2021ctt,Klein:2021les} we will denote by a superscript $(i)$ the gauge where $A=0$ at $r_i$. 

The Unruh vacuum is defined in terms of Unruh-mode solutions. These are classical solutions to the Klein-Gordon equation of the form \eqref{mode ansatz} characterized by means of initial conditions\footnote{The parameter $\xi$ is an arbitrary normalization constant of length dimension which is necessary to have a dimensionless argument inside the logarithm. We choose it consistently for all modes (i.e., all in-modes have the same $\xi$).} on $\mathcal{I}^{-}\cup \mathcal{H}^-\cup \mathcal{H}^\mathcal{L}$

\begin{subequations}\label{unruh modes bc}
\begin{align}
     H_{ \omega \ell}^{\text{in}}&=\begin{cases}
     &\vert \omega \vert^{-1/2}e^{-i\omega v+iqQ \ln{(r_*/\xi)}} \ \hspace{1cm} \text{on } \mathcal{I}^- \\
     & 0 \hspace{4.7cm} \text{on } \mathcal{H}^-\cup \mathcal{H}^\mathcal{L}
     \end{cases}\\
     H_{ \omega \ell}^{(+)\text{up}}&=\begin{cases}
         &0\ \hspace{4.6cm} \text{on } \mathcal{I}^- \\
     &\vert \omega \vert^{-1/2}e^{-i\omega U}  \hspace{2.9cm} \text{on } \mathcal{H}^-\cup \mathcal{H}^\mathcal{L},
     \end{cases}
\end{align}
\end{subequations}
 where the notation $(+)$ denotes that the up-Unruh modes are defined in a gauge where the potential vanishes at $r=r_+$. As usual, it is more convenient to reexpress the Unruh mode solutions in terms of the Boulware modes, which can be separated additionally as \eqref{t separation}. They are defined in regions I and II by their initial data on $\mathcal{H}^{-}\cup \mathcal{I}^-$ and $\mathcal{H}^\mathcal{L}\cup \mathcal{H}^\mathcal{R}$, respectively. The in-II modes have asymptotic behaviour $ h_{\omega \ell}^{(+)\text{in,II}}\sim e^{-i\omega r_*}$ on $\mathcal{H}^\mathcal{R}$ and vanish on $\mathcal{H}^\mathcal{L}$, whereas the up-II modes have asymptotic behaviour $h_{\omega \ell}^{(+)\text{up,II}}\sim e^{i\omega r_*}$ on $\mathcal{H}^\mathcal{L}$ and vanish on $\mathcal{H}^\mathcal{R}$. Similarly, the up-I modes have asymptotic behavior $h_{\omega \ell}^{(+)\text{up,I}}\sim e^{i\omega r_*}$ on $\mathcal{H}^\mathcal{-}$ and vanish on $\mathcal{I^-}$. The in-I modes coincide with the in-Unruh modes in region I.

The scattering coefficients associated to the in-I/II are defined with respect to the following asymptotic behavior
\begin{subequations}
\begin{align}\label{(+)inII modes}
     h_{\omega \ell}^{(+)\text{in,II}}&=\begin{cases}
     &e^{-i\omega r_*} \ \hspace{6.4cm}  r_*\to -\infty \\
     & T_{\omega \ell} ^{\text{in,II}}e^{-i(\omega-\omega_{\text{II}}) r_*}+R_{\omega \ell}^{\text{in,II}}e^{i(\omega-\omega_{\text{II}}) r_*} \hspace{2.1cm} r_*\to \infty
     \end{cases}\\
     h_{\omega \ell}^{\text{in,I}}&=\begin{cases}
     &T_{\omega l} ^{\text{in,I}}e^{-i(\omega-\omega_\text{I}) r_*} \hspace{5.12cm}  r_*\to -\infty \\
     & e^{-i\omega r_*+iqQ\ln(r_*/\xi)}+R_{\omega l}^{\text{in,I}}e^{i\omega r_*-iqQ\ln(r_*/\xi)} \hspace{1.22cm} r_*\to \infty,
     \end{cases}\label{inI modes}
     \end{align}
 \end{subequations}
 with $\omega_\text{I}=\frac{qQ}{r_+}$ and $\omega_{\text{II}}=\frac{qQ}{r_-}-\frac{qQ}{r_+}$. Similarly, the up-I/II modes
  \begin{subequations}
  \begin{align}
      \label{(+)upII modes}
     h_{\omega \ell}^{(+)\text{up,II}}&=\begin{cases}
     &e^{i\omega r_*} \ \hspace{6.6cm}  r_*\to -\infty \\
     & T_{\omega \ell} ^{\text{up,II}}e^{i(\omega-\omega_{\text{II}}) r_*} + R_{\omega \ell} ^{\text{up,II}}e^{-i(\omega-\omega_\text{II}) r_*}\hspace{1.95cm} r_*\to \infty
     \end{cases}\\
\label{(+)upI modes}
     h_{\omega\ell}^{(+)\text{up,I}}&=\begin{cases}
     &e^{i\omega r_*}+ R_{\omega \ell} ^{\text{up,I}}e^{-i\omega r_*} \ \hspace{4.52cm}  r_*\to -\infty \\
     & T_{\omega \ell} ^{\text{up,I}}e^{i(\omega+\omega_{\text{I}}) r_*-iqQ\ln(r_*/\xi)} \hspace{3.6cm} r_*\to \infty.
     \end{cases}
      \end{align}
 \end{subequations}
The scattering problem of the Boulware modes is illustrated in Fig. \ref{fig:scatteringBoulware}. The scattering coefficients satisfy the Wronskian relations
\begin{subequations}\label{wronskian II exact}
    \begin{align}
        1&=\frac{\omega-\omega_\text{II}}{\omega}\big( \vert T_{\omega \ell} ^{\text{up/in,II}}\vert^2-\vert R_{\omega \ell} ^{\text{up/in,II}}\vert^2\big)\\
        1&=\vert R_{\omega \ell} ^{\text{in,I}}\vert^2+\frac{\omega-\omega_\text{I}}{\omega}\vert T_{\omega \ell} ^{\text{in,I}}\vert^2\label{wronskian inI}\\
       1&=\vert R_{\omega \ell} ^{\text{up,I}}\vert^2+\frac{\omega+\omega_\text{I}}{\omega}\vert T_{\omega \ell} ^{\text{up,I}}\vert^2.\label{wronskian upI}
    \end{align}
\end{subequations}
Reexpressing for example the up I/II-modes in terms of the in I/II-modes (and their complex conjugate\footnote{Complex conjugation is not a symmetry of the charged scalar field, but because the radial Klein-Gordon equation \eqref{radial boulware equation} is real, it holds that if $h_{\omega \ell}$ is a solution, so is $\overline{h_{\omega\ell}}$. However, in contrast to the real scalar field, $\overline{h_{\omega\ell}}\neq h_{-\omega\ell}$, see \eqref{t separation}. }) according to their asymptotic behaviour yields relationships
\begin{subequations}\label{relations in up}
\begin{align}
    R_{\omega\ell}^\text{up,II}&=\overline{R_{\omega\ell}^\text{in,II}}\hspace{2.2cm}T_{\omega\ell}^\text{up,II}=\overline{T_{\omega\ell}^\text{in,II}}\\
    T_{\omega-\omega_\text{I}\ell}^\text{up,I}&=\frac{\omega-\omega_\text{I}}{\omega}T_{\omega\ell}^\text{in,I}\hspace{1.5cm}\hspace{-0.5cm}R_{\omega-\omega_\text{I}\ell}^\text{up,I}=-\overline{R_{\omega\ell}^\text{in,I}}\frac{T_{\omega\ell}^\text{in,I}}{\overline{T_{\omega\ell}^\text{in,I}}}.
\end{align}
\end{subequations}
For frequency modes \( \vert\omega\vert \leq \vert\omega_\text{I}\vert \), the prefactor in front of the in/up-I reflection coefficient in the Wronskian \eqref{wronskian II exact} exceeds 1. This effect is known as superradiance, and we refer to the frequency range \( \vert\omega\vert \leq \vert\omega_\text{I}\vert \) as the superradiant band. In RN, superradiance is inherent to the charged scalar field ($\omega_\text{I}$ vanishes in the limit $qQ\to 0$), but in realistic spinning BHs superradiance is a generic feature.

The quantum field operator can be expanded in terms of Unruh mode solutions 
\begin{equation}
    \Phi(x)=\sum_{\Xi, \ell, m}\int_0^\infty d\omega \Big(\Phi_{\omega \ell m}^\Xi (x) a_{\omega \ell m}^\Xi +\Phi_{-\omega \ell m}^\Xi(x) b_{\omega \ell m}^{\Xi\ \dagger} \Big) \hspace{1cm} \Xi\in \{\text{in}, \text{up} \}
\end{equation}
which are appropriately symplectically normalized \cite{Iuliano:2023bqp}. $a_{\omega \ell m}^\Xi$  and $b_{\omega \ell m}^\Xi$ are the standard Fock space annihilation operators satisfying the usual canonical commutation relations. The Unruh state $\ket{0}_\text{U}$ is defined as the ground state of the construction, i.e. the state for which
\begin{equation}
    a_{\omega \ell m}^\Xi \ket{0}_\text{U}=0= b_{\omega \ell m}^\Xi \ket{0}_\text{U}\hspace{0.5cm} \forall\  \omega >0, \ell, m, \hspace{0.5cm} \Xi\in \{\text{in}, \text{up} \}.
\end{equation}
This state is stationary (i.e., invariant under time evolution automorphisms) in region I. In the Reissner-Nordström-de Sitter spacetime, the Unruh state has been demonstrated to be Hadamard in regions I, II, and IV for the real scalar field \cite{Hollands:2019whz} (with region IV extending beyond the cosmological horizon). This result was subsequently generalized to the charged scalar field \cite{Klein:2021les}. A similar proof for the real scalar field was previously established in Schwarzschild \cite{Dappiaggi:2009fx}. To our knowledge, no such proof has been provided for the Reissner-Nordström case, although we anticipate no difficulty in extending the result.
\section{The observables of interest}\label{section: observables}
\subsection{Renormalized mode-sum expressions}
The observables that we will consider are the current density $j_\nu$ and the stress-energy tensor $T_{\mu \nu}$, classically given by
\begin{subequations}\label{observables}
    \begin{align}
        j_\nu&=iq\Big(\Phi(D_\nu\Phi)^*-\Phi^* D_\nu\Phi \Big)\label{classical current}\\
       T_{\mu\nu}&= \frac{1}{2}\Big((D_\mu\Phi )^*D_\nu \Phi + D_\mu\Phi( D_\nu \Phi)^*\Big)-\frac{1}{4}g_{\mu\nu} g^{\rho \lambda}\Big( (D_\rho \Phi )^* D_\lambda \Phi + D_\rho \Phi  (D_\lambda \Phi)^*\Big).     
    \end{align}
\end{subequations}
In particular these expressions are local and quadratic in the field, requiring renormalization in the quantum version. The mode-sum expressions for the renormalized quantum analogs of \eqref{observables} at the horizons are readily available in the literature. Here we shall quote those results. The event horizon (EH) fluxes can be found in \cite{Klein:2021les,Balakumar:2022yvx} to be
\begin{subequations}\label{observables EH}
\begin{align}
   \langle j_v\rangle_\text{U}^{\mathcal{H}^\mathcal{R}}&=\sum_\ell \frac{-q(2\ell+1)}{16\pi^2 r_+^2}\int_0^\infty d\omega \Big(F_\ell (\omega)+F_\ell (-\omega)\Big)\\
  \langle T_{vv}\rangle_\text{U}^{\mathcal{H}^\mathcal{R}}&=\sum_{\ell=0}^\infty\frac{2\ell+1}{32\pi^2 r_+^2}\int_0^\infty d\omega \ \omega \Big( F_\ell(\omega)-F_\ell(-\omega)\Big)\label{RSET eh1} \\ 
     F_\ell (\omega)&=\big(\Theta(\omega+\omega_\text{I})-\Theta (-\omega-\omega_\text{I}) \big)(1-\vert R_{\omega \ell}^{\text{up,I}}\vert^2)+\coth\Big(\frac{\pi \omega}{\kappa_+}\Big)\big(\vert R_{\omega \ell}^{\text{up,I}}\vert^2-1\big),\label{Fls}
\end{align}
\end{subequations}
with $\Theta(x)$ the step-function.\footnote{The expression in \cite{Klein:2021les} for $\langle j_v (x) \rangle_\text{U}$ refers to a conformally massive, charged scalar in a RNdS background. 
Redoing their analysis in the RN case, we find that the RN mode-sums match the $\Lambda\to0$ limit of their expressions.  Equation \eqref{RSET eh1} can be obtained from eq. (4.19b),  (4.21b) in \cite{Balakumar:2022yvx} using regularity of the Unruh state across $\mathcal{H}^\mathcal{R}$.} In order to investigate these expressions further, we assume that $\omega_\text{I}>0$ (i.e. $qQ>0$). The other case is treated analogously. Splitting the integrals and evaluating the step-functions one finds
\begin{subequations}
    \begin{align}
    \langle j_v\rangle_\text{U}^{\mathcal{H}^\mathcal{R}}&=\sum_{\ell=0}^\infty \frac{-q(2\ell+1)}{16\pi^2 r_+^2}\Big\{\int_0^{\omega_\text{I}} d\omega \Big(2-\vert R_{\omega \ell}^{\text{up,I}}\vert^2-\vert R_{-\omega \ell}^{\text{up,I}}\vert^2+\coth\Big(\frac{\pi\omega}{\kappa_+} \Big)(\vert R_{\omega \ell}^{\text{up,I}}\vert^2-\vert R_{-\omega \ell}^{\text{up,I}}\vert^2) \Big)\\ \nonumber
    &\hspace{3cm}+\int_{\omega_\text{I}}^\infty d\omega \Big(\vert R_{\omega \ell}^{\text{up,I}}\vert^2-\vert R_{-\omega \ell}^{\text{up,I}}\vert^2 \Big)\Big( \coth\Big(\frac{\pi\omega}{\kappa_+}\Big)-1\Big)\Big\}\label{current EH} \\
    \langle T_{vv}\rangle_\text{U}^{\mathcal{H}^\mathcal{R}}&=\sum_{\ell=0}^\infty\frac{2\ell+1}{32\pi^2 r_+^2}\Big\{\int_0^{\omega_\text{I}} d\omega \ \omega \Big( \vert R_{-\omega \ell}^{\text{up,I}}\vert^2-\vert R_{\omega \ell}^{\text{up,I}}\vert^2-\coth\Big(\frac{\pi \omega}{\kappa_+} \Big) (2-\vert R_{\omega \ell}^{\text{up,I}}\vert^2-\vert R_{-\omega \ell}^{\text{up,I}}\vert^2)\Big)\\ 
    \nonumber
    &\hspace{3cm}+\int_{\omega_\text{I}}^\infty d\omega \ \omega \Big(2-\vert R_{\omega \ell}^{\text{up,I}}\vert^2-\vert R_{-\omega \ell}^{\text{up,I}}\vert^2)\Big(1-\coth\Big(\frac{\pi \omega}{\kappa_+} \Big) \Big) \Big\}   .\label{stress energy EH}
    \end{align}
\end{subequations}
It is clear that knowledge of the I-scattering coefficients in the range $\vert \omega\vert \in(0, \omega_\text{I})$ is necessary to evaluate the expressions above. The terms involving the scattering coefficients in the $\omega>\omega_\text{I}$ integral are scaled in both cases with an exponentially suppressing factor\footnote{Here and in the following, the symbol "$\gtrsim$" stands for "greater than or of the order of" (and analogously for "$\lesssim$").}
\begin{equation} \label{suppressing}
    \coth\Big(\frac{\pi\omega}{\kappa_+}\Big)-1=\frac{2e^{-\pi\omega/\kappa_+}}{e^{\pi\omega/\kappa_+}-e^{-\pi\omega/\kappa_+}}\ \underset{ \pi\omega\gtrsim \kappa_+}{\xrightarrow{\text{\hspace{8mm}}}}\ 2 e^{-2\pi\omega/\kappa_+}.
\end{equation}
Altogether, this means that knowledge of the  I-scattering coefficients in the frequency range
\begin{equation}\label{frequency range}
    \vert\omega\vert \in (0,\omega_\text{I})\cup \{\vert\omega\vert \lesssim \kappa_+ \}
\end{equation}
suffices for the evaluation of $\langle j_v\rangle_\text{U}^{\mathcal{H}^\mathcal{R}}$ and $\langle T_{vv}\rangle_\text{U}^{\mathcal{H}^\mathcal{R}}$.
In the near-extremal case, where the surface gravities are very small, we simplify our analysis by restricting to small values of $\omega_\text{I}M\ll 1$, corresponding to small field charges $ qQ \ll 1$. Our goal is to develop a series expansion of the observables in both the near-extremality parameter $\Delta$ and the field charge $q$. By considering small charges, we limit the frequency interval required to compute the scattering coefficients, making the expansion in the field charge more meaningful and simplifying the problem.

In the following, we turn to the evaluation of the above observables at the Cauchy horizon. Expressions for the renormalized current density and the renormalized stress energy tensor (SET) at the $\mathcal{CH}^\mathcal{R}$ are available in \cite{ Klein:2021ctt} and can be adapted to the RN case\footnote{In the reference, the $vv$-component of the SET is renormalized by considering the difference of expectation values between the Unruh and another artificial, non-physical state which is constructed to be regular across $\mathcal{CH}^\mathcal{R}$.}, yielding
\begin{subequations}\label{observables IH}
    \begin{align}
        \langle j_v \rangle_\text{U}^{\mathcal{CH}^\mathcal{R}} &= \sum_{\ell=0}^\infty \frac{-q(2\ell+1)}{16\pi^2 (r_-)^2}\int_0^\infty d\omega \Big(G_\ell (\omega)+G_\ell (-\omega) \Big) \label{eq:main_equation} \\
        \langle T_{vv}\rangle_{\text{U}}^{\mathcal{CH}^\mathcal{R}}&=\sum_{\ell=0}^\infty\frac{2\ell+1}{32\pi^2 r_-^2}\int_0^\infty d\omega\  \omega \Big(G_\ell(\omega)-G_\ell(-\omega)-2\coth\Big(\frac{\pi\omega}{\kappa_-} \Big) \Big)
    \end{align}
\end{subequations}
with $G_\ell (\omega) = G_\ell^\text{1} (\omega)+G_\ell^\text{2} (\omega)+G_\ell^\text{3} (\omega)$ given by
\begin{subequations}\label{G functions}
    \begin{align}
        G_\ell^\text{1} (\omega) & = \frac{\omega(\omega^+ +\omega_\text{I})}{(\omega^+)^2} \big( \Theta(\omega^+ +\omega_\text{I})-\Theta(-\omega^+ -\omega_\text{I})\big)\vert T_{\omega^+ \ell}^{\text{up,I}}\vert^2 \vert T_{\omega^+\ell}^{\text{up,II}}\vert^2\label{G1} \\
        G_\ell^\text{2} (\omega) & =\frac{\omega}{\omega^+}\coth\Big(\frac{\pi \omega^+}{\kappa_+} \Big)\Big\{ \vert R_{\omega^+\ell}^{\text{up,II}}\vert^2+\vert R_{\omega ^+\ell}^{\text{up,I}}\vert^2 \vert T_{\omega^+\ell}^{\text{up,II}}\vert^2 \Big\}\label{G2} \\
        G_\ell^\text{3} (\omega) & =\frac{2\omega}{\omega^+}\csch\Big(\frac{\pi \omega^+}{\kappa_+} \Big)\Re \big\{ \overline{R_{\omega^+ \ell}^{\text{up,I}}}T_{\omega^+ \ell}^{\text{up,II}}R_{\omega^+ \ell}^{\text{up,II}} \big\} \label{G3}
    \end{align}
\end{subequations}
and $\omega^+=\omega+\omega_\text{II}$. For notational purposes we also introduce $\omega^-=\omega-\omega_\text{II}$. Splitting the integrals and evaluating the step functions yields 
\begin{equation}\label{current CH}
    \begin{aligned}
        \hspace{-0.5cm}\langle j_v \rangle_\text{U}^{\mathcal{CH}^\mathcal{R}} &= \sum_{\ell=0}^\infty \frac{-q(2\ell+1)}{16\pi^2 (r_-)^2} \Big\{ \int^{\omega_\text{I}+\omega_\text{II}}_0 d\omega\Big[\frac{\omega(\omega^+ +\omega_\text{I})}{(\omega^+)^2}\vert T_{\omega^+\ell}^{\text{up,I}}\vert^2\vert T_{\omega^+\ell}^{\text{up,II}}\vert^2 +\frac{\omega(\omega^--\omega_\text{I})}{(\omega^-)^2}\vert T_{-\omega^-\ell}^{\text{up,I}}\vert^2 \vert T_{-\omega^-\ell}^{\text{up,II}}\vert^2\\
        &\hspace{1cm}+\frac{\omega}{\omega_+}\coth\Big(\frac{\pi\omega^+}{\kappa_+}\Big)\big\{\vert R_{\omega^+\ell}^{\text{up,II}}\vert^2+\vert R_{\omega^+\ell}^{\text{up,I}}\vert^2\vert T_{\omega^+\ell}^{\text{up,II}}\vert^2\big\}+\frac{2\omega}{\omega^+}\csch\Big(\frac{\pi\omega^+}{\kappa_+} \Big)\Re\big\{\overline{R_{\omega^+\ell}^{\text{up,I}}}T_{\omega^+\ell}^{\text{up,II}}R_{\omega^+\ell}^{\text{up,II}}\big\}\\
        &\hspace{1cm}-\frac{\omega}{\omega^-}\coth\Big( \frac{\pi\omega^-}{\kappa_+}\Big)\big\{\vert R_{-\omega^-\ell}^{\text{up,II}}\vert^2+\vert R_{-\omega^-\ell}^{\text{up,I}}\vert^2\vert T_{-\omega^-\ell}^{\text{up,II}}\vert^2\big\} -\frac{2\omega}{\omega^-}\csch\Big( \frac{\pi\omega^-}{\kappa_+}\Big)\Re\big\{ \overline{R_{-\omega^-\ell}^\text{up,I}}T_{-\omega^-\ell}^\text{up,II}R_{-\omega^-\ell}^\text{up,II}\big\}\Big] \\
        &+\int_{\omega_\text{I}+\omega_\text{II}}^\infty\ d\omega\Big[ \frac{\omega(\omega^+ +\omega_\text{I})}{(\omega^+)^2}\vert T_{\omega^+\ell}^{\text{up,I}}\vert^2 \vert T_{\omega^+\ell}^{\text{up,II}}\vert^2\Big( 1-\coth\Big( \frac{\pi \omega^+}{\kappa_+}\Big)\Big)-\frac{\omega(\omega^- -\omega_\text{I})}{(\omega^-)^2}\vert T_{\omega^-\ell}^{\text{up,I}}\vert^2 \vert T_{\omega^-\ell}^{\text{up,II}}\vert^2 \Big( 1-\coth\Big( \frac{\pi \omega^-}{\kappa_+}\Big)\Big)\\
        &\hspace{1cm}+\frac{\omega}{\omega^+}\coth\Big( \frac{\pi \omega^+}{\kappa_+}\Big)\big(\vert R_{\omega^+\ell}^{\text{up,II}}\vert^2+\vert T_{\omega^+\ell}^{\text{up,II}}\vert^2 \big)-\frac{\omega}{\omega^-}\coth\Big(\frac{\pi\omega^-}{\kappa_+} \Big)\big(\vert R_{-\omega^-\ell}^{\text{up,II}}\vert^2+\vert T_{-\omega^-\ell}^{\text{up,II}}\vert^2\big)\\
        &\hspace{1cm}+\frac{2\omega}{\omega^+}\csch\Big(\frac{\pi\omega^+}{\kappa_+}\Big) \Re\big\{\overline{R_{\omega^+\ell}^{\text{up,I}}}T_{\omega^+\ell}^{\text{up,II}}R_{\omega^+\ell}^{\text{up,II}}\big\}-\frac{2\omega}{\omega^-}\csch\Big(\frac{\pi\omega^-}{\kappa_+}\Big)\Re\big\{\overline{R_{-\omega^-\ell}^{\text{up,I}} }T_{-\omega^-\ell}^{\text{up,II}}R_{-\omega^-\ell}^{\text{up,II}}\}\Big]\ \Big\}.
    \end{aligned}
\end{equation}
In the frequency regime $\vert \omega\vert< \omega_\text{I}+\omega_\text{II}$ knowledge of both I- and II-scattering coefficients is required in order to evaluate the integral. In the second integral, the I-scattering coefficients enter in two ways: one of them is scaled by the same\footnote{In the regime $\vert \omega\vert> \omega_\text{I}+\omega_\text{II}$ it is justified to approximate $\omega^+\approx\omega$ as well as $\omega^-\approx \omega$ since corrections are of relative order $\big\vert\frac{\omega^\pm-\omega}{\omega}\big\vert\sim \Delta\ll 1$. This claim follows from the near-extremal assumption and is independent of the field charge $q$.} exponentially suppressing factor as in the event horizon \eqref{suppressing}. The other is multiplied by $\csch\big( \frac{\pi\omega}{\kappa_+}\big)$, which also suppresses contributions from large frequencies (in the sense discussed above at the EH). Therefore, also at the Cauchy horizon, knowledge of the I-scattering coefficients in the frequency range \eqref{frequency range} is sufficient. The II-scattering coefficients additionally enter in a term which is not exponentially suppressed, namely the penultimate line in \eqref{current CH}. However, for large $| \omega |$ the scattering coefficients converge to $1$ and $0$ faster than any power,\footnote{For the scattering at a smooth, exponentially decaying potential, this was shown in \cite{yafaev1992mathematical}. However, the result also holds if the potential exponentially asymptotes to different values at the two sides, as is the case for the potential relevant in the present context.} so that the two terms in the penultimate line of \eqref{current CH} cancel up to a term decaying faster than any power of $\omega$.
Similar arguments hold for the mode-sum expression of $\langle T_{vv}\rangle_{\text{U}}^{\mathcal{CH}^\mathcal{R}}$.

From current and energy conservation in the semiclassical picture (i.e., $\nabla^\mu \big(\langle T_{\mu\nu}\rangle_\Psi+T_{\mu\nu}^\text{EM}\big)=0$), and under the additional assumptions of spherical symmetry and time-translation symmetry, it follows that \cite{Balakumar:2022yvx}
\begin{align}\label{general forms}
  \langle j_{r_*}\rangle_\text{U} & =\frac{\mathcal{K}}{r^2}, &
        \langle T_{tr_*}\rangle_\text{U} & =-\frac{\mathcal{L}}{r^2}+\frac{\mathcal{K}Q}{r^3},
\end{align}
   for some a priori undetermined constants $\mathcal{K}$ and $\mathcal{L}$. The constant $\mathcal{L}$ has, up to a positive factor, the natural interpretation of energy flux at radial infinity. If $\mathcal{L}>0$, then the flux is negative and the black hole is radiating away. If $\mathcal{L}<0$, the energy flux is positive and the black hole is absorbing radiation. Since the Unruh state describes a BH which is evaporating via the emission of Hawking radiation, we expect to find $\mathcal{L}>0$. Similarly, one interprets $\mathcal{K}$ as the flux of charge emitted by the black hole \cite{Balakumar:2022yvx}. If $\mathcal{K}Q>0$, the BH is losing charge and viceversa. Since the Unruh state is regular across $\mathcal{H}^\mathcal{R}$ (and so, by the tensor transformation law, $\langle j_{u}\rangle_\text{U}$, $\langle T_{uu}\rangle_\text{U}$, and $\langle T_{uv}\rangle_\text{U}$ have to vanish there), the constants $\mathcal{K}$ and $\mathcal{L}$ in \eqref{general forms} can be fixed by evaluation of $\langle j_v\rangle _\text{U}$ and $\langle T_{vv}\rangle_\text{U}$ at $\mathcal{H}^\mathcal{R}$.

\section{Scattering coefficients}\label{section: scattering coeffs}
\subsection{Interior region}
The radial equation \eqref{radial KG} can not be solved analytically in closed form the interior region. However, motivated by the results \cite{Zilberman:2021vgz} for the real scalar field, we aim at an approximate solution that allows to extract the scattering coefficients in the near-extremal limit. For this, we expand \eqref{radial KG} in powers of the near-extremality parameter $\Delta$ , defined in \eqref{near extremality parameter}. We will consider the field to be in the $(+)$ gauge, which is most convenient to then extract directly the scattering coefficients in \eqref{(+)inII modes}. The analysis that follows is a generalization of \cite{Zilberman:2021vgz}. We first introduce the rescaled variable
 \begin{equation*}
     s=\frac{r/M-1}{\Delta},
 \end{equation*}
which attains the values $+1$ and $-1$ at the event and inner horizon, respectively. In terms of $s$, the potential $V_\ell$ in \eqref{potential radial KG} can be rewritten exactly as
\begin{equation}
    V_\ell(s)=\frac{\Delta^2}{M^2} \frac{s^2-1}{(1+s\Delta)^4}\Big(\ell(\ell+1)+\frac{2\Delta(s+\Delta)}{(1+s\Delta)^2} \Big)
\end{equation}
and the term coupling to the electromagnetic field is also rewritten exactly as
\begin{equation}
    \Big(\omega-qQ\Big(\frac{1}{r}-\frac{1}{r_+}\Big) \Big)^2=\Big(\omega-qQ\frac{\Delta}{M}\frac{1-s}{(1+\Delta)(1+\Delta s)}\Big)^2.
\end{equation}
The variable $s$ is related to the tortoise coordinate by
\begin{equation*}
    \frac{ds}{dr_*}=\frac{\Delta(s^2-1)}{M(1+\Delta s)^2}.
\end{equation*}
It is most useful to consider the rescaled, dimensionless quantities $\Tilde{r}_*\equiv(\Delta/M)r_*$ and $\Tilde{\omega}\equiv (M /\Delta) \omega$. In terms of these, 
$
     ds/d\Tilde{r}_*=s^2-1+\mathcal{O}(\Delta)
$, which can be integrated to leading order in $\Delta$, i.e.
$s(\Tilde{r}_*)=-\tanh{\Tilde{r}_*}+\mathcal{O}(\Delta)$.

\begin{figure}[h!]
    \centering
    \begin{subfigure}{0.49\textwidth}
        \centering
        \includegraphics[width=\linewidth]{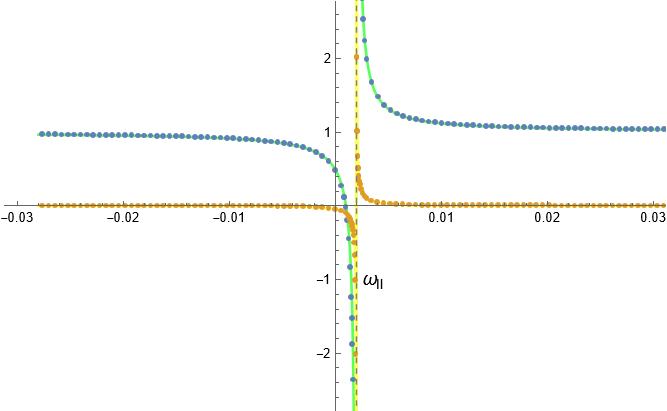}
        \caption{$T_{\omega \ell=0}^{\text{in,II}}$}
    \end{subfigure}
    \hfill
    \begin{subfigure}{0.49\textwidth}
        \centering
        \includegraphics[width=\linewidth]{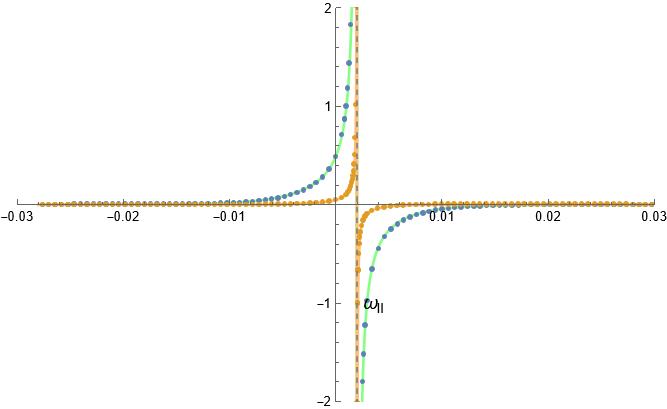}
        \caption{$R_{\omega \ell=0}^{\text{in,II}}$}
    \end{subfigure}
    
    \medskip
    
    \begin{subfigure}{0.49\textwidth}
        \centering
        \includegraphics[width=\linewidth]{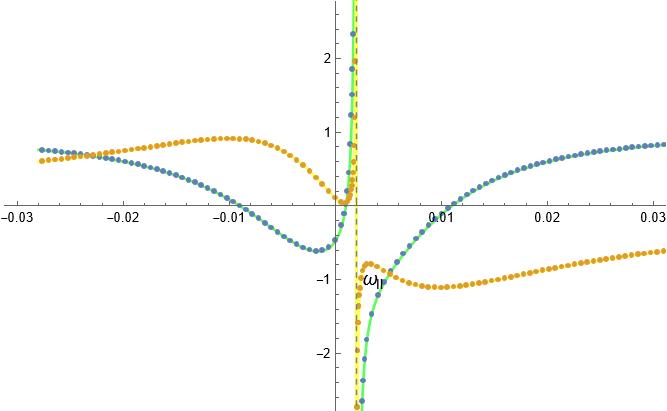}
        \caption{$T_{\omega \ell=1}^{\text{in,II}}$}
    \end{subfigure}
    \hfill
    \begin{subfigure}{0.49\textwidth}
        \centering
        \includegraphics[width=\linewidth]{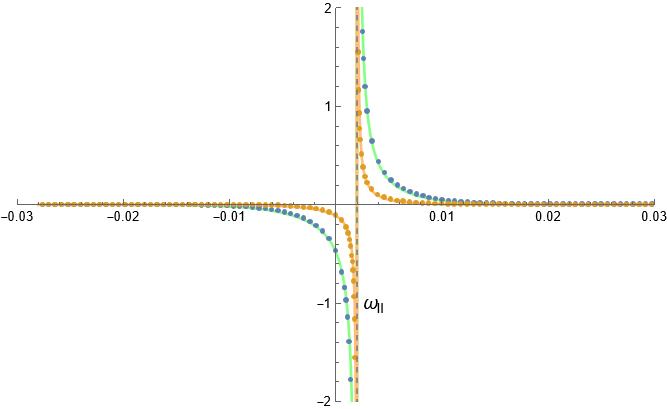}
        \caption{$R_{\omega \ell=1}^{\text{in,II}}$}
    \end{subfigure}
    
    \medskip
    
    \begin{subfigure}{0.49\textwidth}
        \centering
        \includegraphics[width=\linewidth]{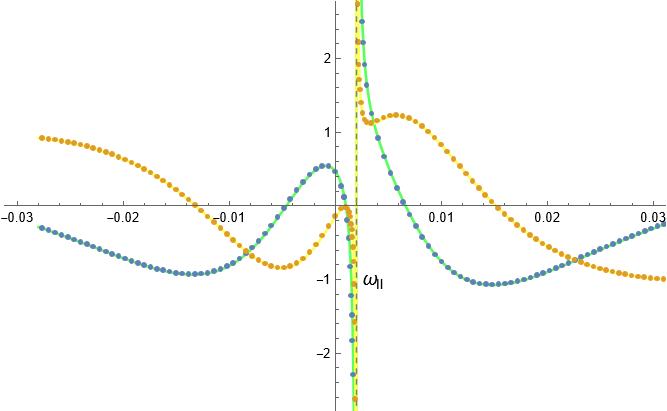}
        \caption{$T_{\omega \ell=2}^{\text{in,II}}$}
    \end{subfigure}
    \hfill
    \begin{subfigure}{0.49\textwidth}
        \centering
        \includegraphics[width=\linewidth]{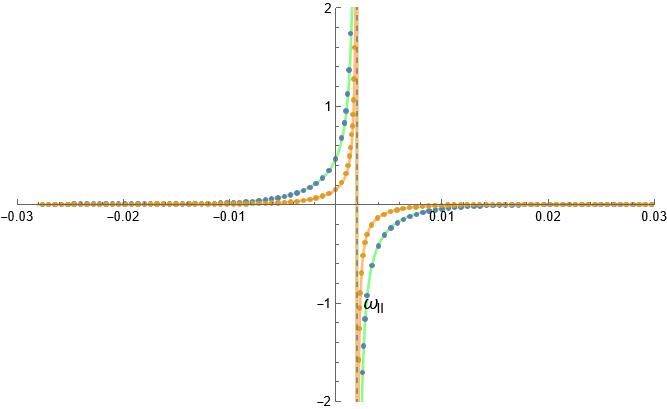}
        \caption{$R_{\omega \ell=2}^{\text{in,II}}$}
    \end{subfigure}
    
    \caption{Comparison between analytical \eqref{sc coeff II} and numerically obtained real (blue) and imaginary (orange) parts of the scattering coefficients in the interior region II, for the parameter values $\Delta=10^{-2}$, $M=1$, $qQ=10^{-1}$, and $\omega_\text{II}\approx 0.002$. The numerical results are obtained by bringing the radial Klein-Gordon equation \eqref{radial boulware equation} into the form of the confluent Heun equation, solutions of which are implemented in Mathematica \cite{Mathematica}.}
    \label{fig:sc coeff II}
\end{figure}

Finally, combining the above and rewriting \eqref{radial KG} explicitly in terms of $\Tilde{r}_*$ to linear order in the near-extremality parameter yields the equation
\begin{equation}
    \begin{aligned}\label{general II radial equation}
    \partial^2_{\Tilde{r}_*}h_{\omega \ell}^{(+)\text{in,II}}=\Big\{-\ell(\ell+1)\sech^2{(\Tilde{r}_*)} -\big( \Tilde{\omega}-qQ(1+\tanh{(\Tilde{r}_*)})\big)^2+\mathcal{O}\big( \Delta\big)\Big\}h_{\omega \ell}^{(+)\text{in,II}},
    \end{aligned}
\end{equation}
which can be solved analytically in closed form in terms of hyperbolic and hypergeometric functions (see Appendix \ref{AppendixA} for the details). We find
\begin{subequations}\label{sc coeff II}
\begin{align}
 &\hspace{-1.1cm}T_{\omega \ell} ^{\text{in,II}}=\frac{\pi\Tilde{\omega}\csch{\big(\pi(\Tilde{\omega}-2qQ)\big)}\Gamma(-i\Tilde{\omega})}{\Gamma\big(1+i(\Tilde{\omega}-2qQ)\big)\Gamma\big(\frac{1+\alpha}{2}-i(\Tilde{\omega}-2qQ) \big)\Gamma\big(\frac{1-\alpha}{2}-i\Tilde{\omega} \big)} +\mathcal{O}(\Delta)\label{T in II}\\
R_{\omega \ell} ^{\text{in,II}}=&\frac{\Tilde{\omega}\Big(\cosh\big(2\pi qQ\big)+\cosh\big( 2\pi qQ-i\pi \alpha\big) \Big)}{2\pi \Gamma\big(1-i(\Tilde{\omega}-2qQ) \big)\sinh\big(\pi(2qQ-\Tilde{\omega})\big)} \Gamma(-i\Tilde{\omega})\Gamma
\big( \scalebox{0.95}{\( \frac{1-\alpha}{2}\)}
\big)\Gamma
\big( \scalebox{0.95}{\( \frac{1+\alpha}{2}+2iqQ\)}
\big)+\mathcal{O}(\Delta)\label{R in II},
\end{align}
\end{subequations}
with the constant $\alpha$ defined in \eqref{alpha}. See Fig. \ref{fig:sc coeff II} for comparison with numerical results. The reflection coefficient \eqref{R in II} is exponentially suppressed at large frequencies, with scale set by the surface gravity, that is, $\sim e^{-\frac{\omega}{\Delta/M}} $.
The analog of equation \eqref{general II radial equation} is also analytically solvable if one considers a massive field. This is implemented by substituting $\ell(\ell+1)\longmapsto \ell(\ell+1)+m^2$. In the limit $qQ\to 0$ the potential in \eqref{general II radial equation} reduces to a potential well of modified Pöschl-Teller type \cite{flugge1999practical}, for which the reflection coefficient is known to vanish if $\ell$ is an integer (agreeing with the results for the uncharged scalar \cite{Zilberman:2021vgz}). In this case, the divergence at $\omega_\text{II}$ in Fig. \ref{fig:sc coeff II} is absent. A similar potential with an identically vanishing reflection coefficient also occurs in the context of quantization of perturbations in a kink background \cite{Martin:2022pri}. 
    
\subsection{Exterior region}
We solve equation \eqref{radial KG} in the exterior region I, in the near-extremal domain $\Delta\ll1$ for small field charges $\vert qQ\vert \ll 1$. Previously we argued that it is sufficient to solve the scattering problem in the superradiant frequency regime $\vert\omega\vert \in (0,\omega_I)$, and up to the order of the surface gravities, see \eqref{frequency range}. To this end, we consider incoming wave solutions from the event horizon of the form
\begin{equation}\label{BC practice}
     \psi_{\omega \ell}=\begin{cases}
     &e^{-i(\omega-\omega_\text{I}) r_*}   \hspace{6.75cm}  r_*\to -\infty \\
     & C_{\omega\ell}e^{-i\omega r_*+iqQ\ln(r/\xi)}+D_{\omega\ell}e^{i\omega r_*-iqQ\ln(r/\xi)} \hspace{1.92cm}r_*\to \infty.
     \end{cases}
 \end{equation}
The scattering coefficients \eqref{inI modes} are recovered from the factors $C_{\omega\ell}$ and $D_{\omega\ell}$ by
\begin{equation}\label{relation cd}
    T_{\omega\ell}^\text{in,I}=\frac{1}{C_{\omega\ell}}\hspace{1cm} R_{\omega\ell}^\text{in,I}=\frac{D_{\omega\ell}}{C_{\omega\ell}}.
\end{equation}
The idea how to approach this problem was first developed in \cite{Zilberman:2021vgz} and involves partitioning the exterior region into overlapping subregions where different terms in the radial equation \eqref{radial KG} can be neglected. One finds approximate solutions in the different subregions that can be matched in the overlaps. From these one extracts the coefficients $C_{\omega\ell}$ and $D_{\omega\ell}$ at radial infinity (to leading order in the small parameters $\Delta$ and $qQ$). For technical reasons, we will allow for generic values of $L^2\equiv\ell(\ell+1)$ (in particular non-integer ones). This is a standard strategy when matching analytical asymptotic expansions \cite{Page:1976df}. 
We find the results are analytic in $L^2$, allowing to reintroduce integral $\ell$ values in the final stage of the analysis.

\begin{center}
    \textbf{Region A: vicinity of the EH}
\end{center}
The region A is characterized by
\begin{equation}\label{regionI}
    \frac{r-r_+}{M}\ll 1 \hspace{2cm} \text{(region A)}. 
\end{equation}
We define two additional small parameters 
\begin{equation}
    \epsilon\equiv \frac{r-r_+}{M}, \hspace{1cm} \delta\equiv \vert qQ\vert,
\end{equation}
quantifying proximity to the event horizon and the smallness of the field charge. It follows that $\frac{1}{r_+}=\frac{1}{M}+\mathcal{O}\big( \frac{\Delta}{M}\big)$ and $  \frac{1}{r}=\frac{1}{r_+}+\mathcal{O}\big(\frac{\epsilon}{M} \big)$. Other terms in the radial equation are estimated as
\begin{subequations}\label{f(r) region A}
\begin{align}
     f(r)\big\vert_{\text{region A}}&=\frac{r-r_+}{M}\Big( \frac{r-r_+}{M}+2\Delta\Big) +\mathcal{O}(\epsilon \Delta^2, \epsilon^2\Delta,\epsilon^3 )\label{f(r) region A sub}\\
      \frac{\partial_r f(r)}{r}\big\vert_{\text{region A}}&=\mathcal{O}\Big(\frac{\epsilon}{M^2}, \frac{\Delta}{M^2} \Big)\\
      \Big(\omega-\frac{qQ}{r}\Big)^2\bigg\vert_{\text{region A}}&= (\omega-\omega_\text{I})^2+\mathcal{O}\Big(\frac{\epsilon \delta^2}{M^2} \Big),
\end{align}
\end{subequations}
with $\omega_\text{I}=qQ/r_+$.  
    Additionally, the tortoise coordinate $r_*(r)\equiv r+\frac{1}{2\kappa_+}\log\Big( \frac{r-r_+}{M}\Big)-\frac{1}{2\kappa_-}\log\Big(\frac{r-r_-}{M} \Big)$ behaves at leading order in $\Delta$ like\footnote{In practice, this $r_*$ approximation neglects the linear term in $r$ (justified since $\vert r_*\vert \gg M$ very close to the EH) and approximates the surface gravities \eqref{surface gravity} by their leading term. Note that both surface gravities coincide to leading order in the extremality parameter, so that they are indistinguishable within our approximation. }
     \begin{equation}
        2\kappa_+ r_*(r)=-\log\Big(1+\frac{2M\Delta}{r-r_+}\Big) +\mathcal{O}\Big(\Delta, \Delta\log\Delta,\Delta \epsilon M \Big).\label{tortoise region A}
    \end{equation} Altogether, the radial equation in region A can be rewritten as 
\begin{equation}\label{eq regionI}
\begin{aligned}
    \frac{d\psi_{\omega\ell}^{\text{A}}}{dr_*^2}&=\Big\{\frac{r-r_+}{M}\Big( \frac{r-r_+}{M}+2\Delta\Big)\frac{L^2}{M^2} -(\omega -\omega_\text{I} )^2+M^{-2}\mathcal{O}\Big(\Delta\epsilon^2,\Delta^2\epsilon,\epsilon^3,\epsilon\delta^2 \Big)\Big\}\psi_{\omega\ell}^{\text{A}}\\
     &=\Big\{\frac{L^2\Delta^2}{M^2}\csch^2 (r_*\kappa_+)-(\omega -\omega_\text{I})^2+M^{-2}\mathcal{O}\Big(\Delta\epsilon^2,\Delta^2\epsilon,\epsilon^3,\epsilon\delta^2 \Big)\Big\}\psi_{\omega\ell}^{\text{A}},
\end{aligned}
\end{equation}
and has an analytic solution involving hypergeometric functions
\begin{equation}
\begin{aligned}
    \psi_{\omega\ell}^{\text{A}}&=a_1\ e^{-i(\omega -\omega_\text{I})r_*}\big(-1+e^{2\kappa_+ r_*}\big)^{\frac{1}{2}+
\frac{\sqrt{4L^2+1}}{2}}\tensor[_{2}]{F}{_{1}}\Big[\mathlarger{{\scriptstyle\frac{1}{2}+\frac{\sqrt{4L^2+1}}{2}}},\mathlarger{{\scriptstyle\frac{1}{2}+\frac{\sqrt{4L^2+1}}{2}}}-i\mathlarger{\scriptstyle\big(\frac{\omega-\omega_\text{I}}{\kappa_+}\big)},1-i\mathlarger{\scriptstyle\big(\frac{\omega-\omega_\text{I}}{\kappa_+}\big)},e^{2r_*\kappa_+}\Big]\\
    &+a_2\  e^{-\pi\frac{\omega-\omega_I}{\kappa_+}}e^{i(\omega -\omega_\text{I}) r_*}\big(-1+e^{2\kappa_+ r_*}\big)^{\frac{1}{2}+\frac{\sqrt{4L^2+1}}{2}}\tensor[_{2}]{F}{_{1}}\Big[\mathlarger{{\scriptstyle\frac{1}{2}+\frac{\sqrt{4L^2+1}}{2}}},\mathlarger{{\scriptstyle\frac{1}{2}+\frac{\sqrt{4L^2+1}}{2}}}+i\mathlarger{\scriptstyle\frac{\omega-\omega_\text{I}}{\kappa_+}},1+i\mathlarger{\scriptstyle\frac{\omega-\omega_\text{I}}{\kappa_+}},e^{2r_*\kappa_+}\Big].
\end{aligned}
\end{equation}
In order to fix the constants $a_1$ and $a_2$ one carries the above solution to $r\to r_+$ (or $r_*\to -\infty)$ and imposes the boundary condition \eqref{BC practice}. In practice, consider the limit of the hypergeometric functions when $r\to r_+$ (or, equivalently, $e^{2r_*\kappa_+}\to 0$). Noting that
 $\tensor[_{2}]{F}{_{1}}[a,b,c,\vert x\vert \ll 1]= 1 +\mathcal{O}(x)$, one easily finds 
\begin{equation}
    a_1= (-1)^{\mathlarger{{\scriptstyle\frac{1}{2}+\frac{\sqrt{4L^2+1}}{2}}}}, \hspace{2cm} a_2=0.
\end{equation}
We have therefore determined the approximate solution 
\begin{equation}\label{sol regionI}
     \psi_{\omega\ell}^{\text{A}}=a_1\ e^{-i(\omega -\omega_\text{I}) r_*}\big(-1+e^{2\kappa_+ r_*}\big)^{\mathlarger{{\scriptstyle\frac{1}{2}+\frac{\sqrt{4L^2+1}}{2}}}}\tensor[_{2}]{F}{_{1}}\Big[\mathlarger{{\scriptstyle\frac{1}{2}+\frac{\sqrt{4L^2+1}}{2}}},\mathlarger{{\scriptstyle\frac{1}{2}+\frac{\sqrt{4L^2+1}}{2}}}-i\mathlarger{\scriptstyle\big(\frac{\omega-\omega_\text{I}}{\kappa_+}\big)},1-i\mathlarger{\scriptstyle\big(\frac{\omega-\omega_\text{I}}{\kappa_+}\big)},e^{2r_*\kappa_+}\Big].
\end{equation}
as the approximate solution\footnote{In particular, setting $L=0$ (i.e. $\ell=0$) yields a free wave $
     \psi_{\omega\ell=0}^{\text{A}}=e^{-i(\omega -\omega_\text{I}) r_*}.$ } throughout region A. 

     \begin{center}
    \textbf{Region B: intermediate region}
\end{center}
We would like this region to be characterized by 
\begin{equation}\label{condition ii potential}
    V_{\ell}(r)\gg (\omega -\omega_\text{I})^2 \hspace{2cm}\text{(region B)},
\end{equation}
for which we know that an approximate analytic solution to the equation exists (the \textit{static solution} \cite{Zilberman:2021vgz}). The near-EH behavior of the potential is dominated by the term $\propto L^2$. Therefore, for generic values\footnote{Strictly speaking, one should distinguish very small values of $L^2$. However, we found that the distinction does not have an effect on the final results for the scattering coefficients. } of $L$, criterion \eqref{condition ii potential} translates into $\frac{r-r_+}{M}\gg \vert \omega -\omega_\text{I}\vert M$. On the other hand, for large $r$, the potential in the radial equation \eqref{potential radial KG} decays like
$V_\ell(r)\propto L^2/r^2$. Therefore, condition \eqref{condition ii potential} holds as long as $r^{-2}\gg(\omega-\omega_I)^2
$. Combining these, we reformulate condition \eqref{condition ii potential} as
\begin{subequations}
\begin{align}
    \frac{r-r_+}{M}&\gg \vert \omega -\omega_\text{I}\vert M \hspace{1cm}\text{(small-$r$ limit of region B)}\label{rcondition ii r coordinate}\\
    \frac{r}{M}&\ll \frac{1}{\vert \omega -\omega_\text{I}\vert M}\hspace{1cm} \text{(large-$r$ limit of region B).}\label{large r condition ii}
\end{align}
\end{subequations}
In region B, the radial equation takes the form
\begin{equation}\label{eq region Ai}
f(r)\Big(f'(r)\frac{d\psi_{\omega\ell}^{\text{B}}}{dr} +f(r)\frac{d^2\psi_{\omega\ell}^{\text{B}}}{dr^2}\Big)=\Big\{ f(r)\Big(\frac{L^2}{r^2} +\frac{\partial_r f(r)}{r}\Big)+M^{-2}\mathcal{O}\big( \Delta^2, \delta^2, \Delta\delta \big)\Big\}\psi_{\omega\ell}^{\text{B}}.
\end{equation}
Approximating $f(r) =\frac{(r-r_+)^2}{r^2}\big(1+\mathcal{O}(\Delta) \big)$, yields the solution 
\begin{equation}\label{solutionII}
    \psi_{\omega\ell}^{\text{B}}=b_1\ r (r-r_+)^{-\frac{1}{2}-\frac{\sqrt{4L^2+1}}{2}}+b_2\ r (r-r_+)^{-\frac{1}{2}+\frac{\sqrt{4L^2+1}}{2}},
\end{equation}
where $b_1$ and $b_2$ are two constants to be fixed by matching this solution to the one in region A. This can only be done if both regions overlap, which is the case provided that $\vert \omega-\omega_\text{I}\vert$ is sufficiently small 
\begin{equation}\label{overlap i-ii}
    \vert \omega-\omega_\text{I}\vert M\ll \frac{r-r_+}{M}\ll 1 \hspace{1cm}\text{(overlap regions A-B)}.
\end{equation}
In practice, we apply the left inequality to the solution in region A and the right one to the solution in region  B. Starting with the solution in region B, restriction to the overlap results in\footnote{For the purposes of analyzing the leading behaviour asymptotics, it is equivalent to exchange $r_+\longleftrightarrow M$ like we do in the following equation (since they coincide to leading order in the near-extremal extremal regime). }
\begin{equation}\label{solII overlap I-II}
\psi_{\omega\ell>0}^\text{B}\Big\vert_{\text{overlap A-B}}=b_1\ M (r-r_+)^{-\frac{1}{2}-\frac{\sqrt{4L^2+1}}{2}}+b_2\ M (r-r_+)^{-\frac{1}{2}+\frac{\sqrt{4L^2+1}}{2}}.
\end{equation}
In order to determine the approximate behavior of solution A in the overlap, we look at its asymptotics in the outermost regime of region A, further from the event horizon; see left hand side of \eqref{overlap i-ii}. It then follows that $\vert( \omega-\omega_\text{I})r_*\vert \ll 1$ (this can be proven by taking \eqref{tortoise region A} as a definition for $r_*$ and restricting it to the larger-$r$ regime of A, where ${\scriptstyle \frac{\scriptstyle M\Delta}{\scriptstyle r-r_+}}\ll 1$).

In particular, this allows to expand the exponential in \eqref{sol regionI} to leading order as a Taylor series. Inserting this into \eqref{sol regionI} and expanding the hypergeometric function in a series around $\vert r_* \kappa_+\vert \ll 1$ with $\textit{Mathematica}$ \cite{Mathematica}, one finds the approximate behaviour
\begin{equation}\label{solI overlap I-II}
\begin{aligned}
\psi_{\omega\ell}^{\text{A}}\Big\vert_{\text{overlap A-B}}= a_1& \big(-1+e^{2\kappa_+r_*} \big)^{\frac{1}{2}+\frac{\sqrt{4L^2+1}}{2}}\Big\{ \lambda_1+\mathcal{O}\big(1-e^{2\kappa_+ r_*} \big)\Big\}\\
&+a_1 \big(-1+e^{2\kappa_+r_*} \big)^{\frac{1}{2}-\frac{\sqrt{4L^2+1}}{2}}\Big\{ \lambda_2+\mathcal{O}\big(1-e^{2\kappa_+ r_*} \big)\Big\},
\end{aligned}
\end{equation}
where we introduced the constants 
\begin{subequations}
\begin{align}
    \lambda_1&=-\frac{\pi \csc\big( \sqrt{4L^2+1}\pi \big)\Gamma\big( 1-i\frac{\omega-\omega_\text{I}}{\kappa_+}\big)}{\Gamma\big(\frac{1}{2}-\frac{\sqrt{4L^2+1}}{2} \big)\Gamma\big( 1+\sqrt{4L^2+1}\big)\Gamma\big( \frac{1}{2}-\frac{\sqrt{4L^2+1}}{2}-i\frac{\omega-\omega_\text{I}}{\kappa_+}\big)}\\
    \lambda_2&=\frac{\pi \csc\big( \sqrt{4L^2+1}\pi \big)\Gamma\big( 1-i\frac{\omega-\omega_\text{I}}{\kappa_+}\big)}{\Gamma\big(\frac{1}{2}+\frac{\sqrt{4L^2+1}}{2} \big)\Gamma\big( 1-\sqrt{4L^2+1}\big)\Gamma\big( \frac{1}{2}+\frac{\sqrt{4L^2+1}}{2}-i\frac{\omega-\omega_\text{I}}{\kappa_+}\big)}.
\end{align}
\end{subequations}
 Moreover, from \eqref{tortoise region A} it follows that $ e^{2\kappa_+ r_*}-1=-\frac{2M\Delta}{r-r_+}  \Big( 1+\mathcal{O}\big({\scriptstyle \frac{\scriptstyle M\Delta}{\scriptstyle r-r_+}}\ll 1\big)\Big)$ in the outermost part of region A. Using this, we rewrite \eqref{solI overlap I-II} as
\begin{equation}\label{fin sol B}
    \begin{aligned}\psi_{\omega\ell}^{\text{A}}\Big\vert_{\text{overlap A-B}}= & \Big(\frac{r-r_+}{2M\Delta} \Big)^{-\frac{1}{2}-\frac{\sqrt{4L^2+1}}{2}}\Big\{ \lambda_1+\mathcal{O}\big({\scriptstyle \frac{\scriptstyle M\Delta}{\scriptstyle r-r_+}}\ll 1\big) \Big\}+\Big(\frac{r-r_+}{2M\Delta} \Big)^{-\frac{1}{2}+\frac{\sqrt{4L^2+1}}{2}}\Big\{ \lambda_2+\mathcal{O}\big({\scriptstyle \frac{\scriptstyle M\Delta}{\scriptstyle r-r_+}}\ll 1\big) \Big\},
    \end{aligned}
\end{equation}
which can now be easily matched to the solution in region B. For generic values of $L^2$, we can interpret the two solutions in \eqref{solI overlap I-II} as two independent Fröbenius solutions around $\vert r_* \kappa_+\vert \approx 0$. To match the two solutions means to identify the respective coefficients in \eqref{solII overlap I-II} and \eqref{fin sol B} as
\begin{equation}
    b_1=\lambda_1(2\Delta)^{\frac{1}{2}+\frac{\sqrt{4L^2+1}}{2}}M^{^-\frac{1}{2}+\frac{\sqrt{4L^2+1}}{2}}, \hspace{1cm} b_2=\lambda_2 (2\Delta)^{\frac{1}{2}-\frac{\sqrt{4L^2+1}}{2}} M^{-\frac{1}{2}-\frac{\sqrt{4L^2+1}}{2}}.
\end{equation}

\begin{figure}[h!]
    \centering
    \begin{subfigure}{0.49\textwidth} 
        \centering
        \includegraphics[width=\linewidth]{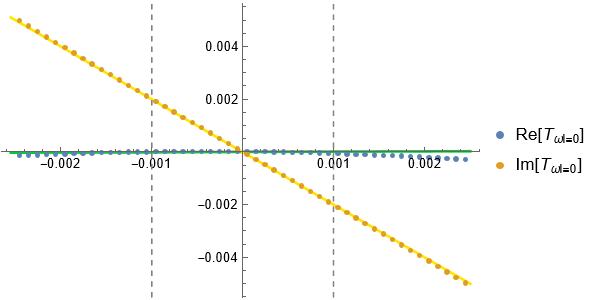}
        \caption{$T_{\omega \ell=0}^{\text{up,I}}$}
    \end{subfigure}
    \hfill
    \begin{subfigure}{0.49\textwidth} 
        \centering
        \includegraphics[width=\linewidth]{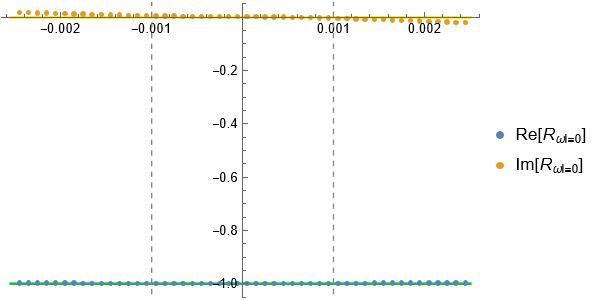}
        \caption{$R_{\omega \ell=0}^{\text{up,I}}$}
    \end{subfigure}
    
    \medskip
    
    \begin{subfigure}{0.49\textwidth} 
        \centering
        \includegraphics[width=\linewidth]{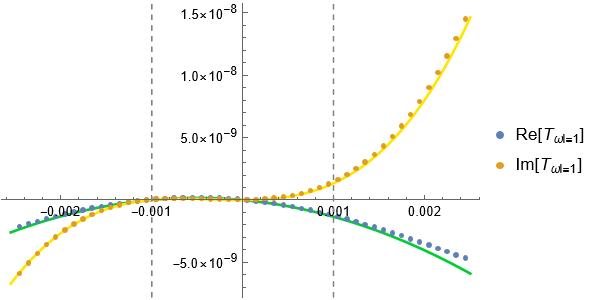}
        \caption{$T_{\omega \ell=1}^{\text{up,I}}$}
    \end{subfigure}
    \hfill
    \begin{subfigure}{0.49\textwidth} 
        \centering
        \includegraphics[width=\linewidth]{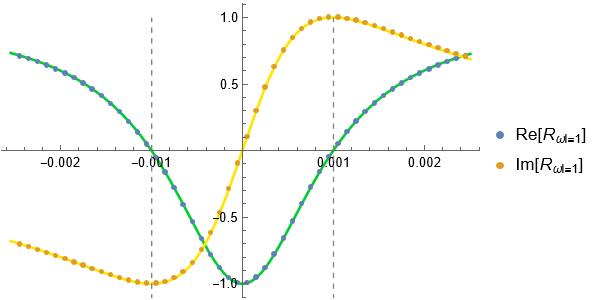}
        \caption{$R_{\omega \ell=1}^{\text{up,I}}$}
    \end{subfigure}
    
    \medskip
    
    \begin{subfigure}{0.49\textwidth} 
        \centering
        \includegraphics[width=\linewidth]{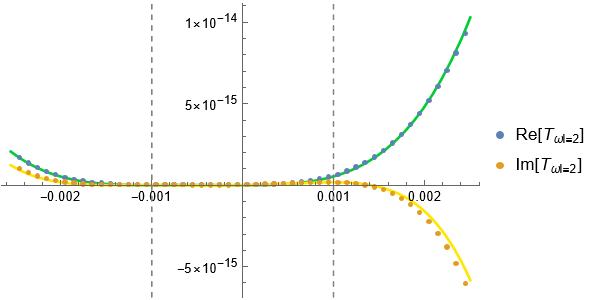}
        \caption{$T_{\omega \ell=2}^{\text{up,I}}$}
    \end{subfigure}
    \hfill
    \begin{subfigure}{0.49\textwidth} 
        \centering
        \includegraphics[width=\linewidth]{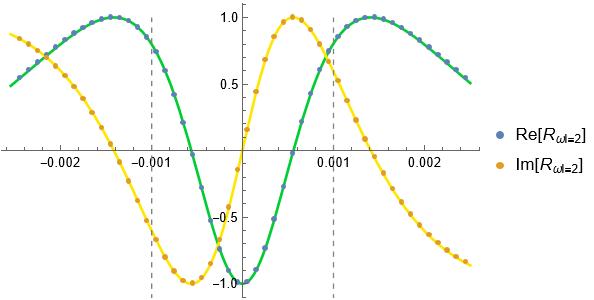}
        \caption{$R_{\omega \ell=2}^{\text{up,I}}$}
    \end{subfigure}
    
    \caption{Comparison between analytical \eqref{sc exterior} and numerical scattering coefficients in the exterior region I for
    parameter values $\Delta=0.001$, $M=1$, $qQ=0.001$. The dashed lines indicate the superradiant frequency $\pm \omega_\text{I}$.}
    \label{fig:sccoi}
\end{figure}

\begin{center}
    \textbf{Region C: quasi-flat region}
\end{center}
  We define the next region by
\begin{subequations}
\begin{align}
   & \frac{r}{M}\gg 1 \hspace{3cm} \text{(small-$r$ limit of region C),}\label{region Aii}\\
 & \log(r/M)\lesssim\frac{1}{(\omega M)(qQ)}\hspace{0.55cm}\text{(large-$r$ limit of region C)}.\label{large r C}
    \end{align}
\end{subequations}
which overlaps with region B if $\vert \omega-\omega_\text{I}\vert M$ is sufficiently small, see \eqref{large r condition ii}. In this region, one has $f(r)=1-2M/r+\mathcal{O}\big(M^2/r^2 \big),$ and $ \partial_r f(r)/r=\mathcal{O}\big( M/r^3\big)$. Collecting the terms up to $r^{-2}$ and neglecting higher orders, we rewrite the radial equation as
\begin{equation}\label{eq iii a}
    \frac{d^2\psi_{\omega \ell>0}^{\text{C}}}{dr_*^2}=\Big\{\frac{L^2}{r^2}-\Big(\omega-\frac{qQ}{r} \Big)^2+\mathcal{O}\Big(\frac{M}{r^3} \Big) \Big\}\psi_{\omega \ell>0}^{\text{C}}.
\end{equation}
Since additionally $qQ\ll 1$, we will also neglect the term quadratic in the field charge
\begin{equation}\label{eq iii b}
        \frac{d^2\psi_{\omega \ell}^{\text{C}}}{dr_*^2}=\Big\{\frac{L^2}{r^2}-\omega^2+2\omega\frac{qQ}{r} +\mathcal{O}\Big(\frac{M}{r^3}, \frac{q^2Q^2}{r^2} \Big) \Big\}\psi_{\omega \ell}^{\text{C}}.
\end{equation}
From the large-$r$ form of $f(r)$ it follows that $r$ and $r_*$ have the same leading behaviour, up to a relative correction that scales as $\propto (M/r)\log(r/M)$. Substituting this in \eqref{eq iii b} and expanding around $r_*\gg (r_*-r)\sim M\log(r/M)$ yields
\begin{equation}
    \begin{aligned}\label{eq region Aii b}
        \frac{d^2\psi_{\omega \ell>0}^{\text{C}}}{dr_*^2}=\Big\{\frac{L^2}{r_*^2}-\omega^2+2\omega\frac{qQ}{r_*}+\mathcal{O}\Big( \frac{L^2 M\log(r/M)}{r^3},\frac{qQ\omega M\log(r/M)}{r^2},\frac{q^2Q^2}{r^2}\Big)\Big\}\psi_{\omega \ell>0}^{\text{C}}.
    \end{aligned}
\end{equation}
For arbitrarily large values of $r$ (or $r_*$), correction terms in \eqref{eq region Aii b} are not negligible anymore. The cut off 
 \eqref{large r C} is chosen accordingly. 
The solution to \eqref{eq region Aii b} is a linear combination of \textit{Whittaker functions}
\begin{equation}\label{solution iii}
\begin{aligned}
    \psi_{\omega \ell}^{\text{C}}&=c_1M_{iqQ, \sqrt{1+4L^2}/2}(2ir_*\omega)+c_2 W_{iqQ, \sqrt{1+4L^2}/2}(2ir_*\omega),
\end{aligned}
\end{equation}
which are classified in \cite{Mathematica} as \textit{WhittakerM} and \textit{WhittakerW}. In order to fix the constants $c_1$ and $c_2$ we carry the above solutions to the overlap region
\begin{equation}\label{overlap II-III}
    1\ll \frac{r}{M}\ll \frac{1}{\vert \omega -\omega_I\vert M}\hspace{0.7cm} \text{(overlap B-C)}
\end{equation}
and match it to the solution from region B
\begin{equation}\label{sol ii overlap ii-iii}
    \psi_{\omega l>0}^{\text{B}}\Big\vert_{\text{overlap B-C}}=b_1 r^{\frac{1}{2}-\frac{\sqrt{4L^2+1}}{2}}+b_2 r^{\frac{1}{2}+\frac{\sqrt{4L^2+1}}{2}}.
\end{equation}
From \eqref{overlap II-III} it follows that we are interested in the small-$ r$ asymptotics of the Whittaker functions. This is implemented by expanding each of the linearly independent solutions around $\vert \omega r\vert \sim \vert \omega r_* \vert \ll 1$\footnote{Everywhere in region C it holds that $r$ and $r_*$ have the same leading behaviour, i.e. $\mathcal{O}(\omega r)=\mathcal{O}(\omega r_*)$, in particular also in its overlap with region B.}, see $13.5$ in \cite{abramowitz1948handbook}

\begin{subequations}\label{TK expansions}
\begin{align}
    M_{iqQ, \sqrt{4L^2+1}/2}(2ir_*\omega)\Big\vert_{\text{overlap B-C}}&=(2i\omega r_*)^{\frac{1}{2}+\frac{\sqrt{4L^2+1}}{2}}\big( 1+\mathcal{O}(\omega r_*)\big)\label{M}\\
     W_{iqQ, \sqrt{4L^2+1}/2}(2ir_*\omega)\Big\vert_{\text{overlap B-C}}&=(2i\omega r_*)^{\frac{1}{2}-\frac{\sqrt{4L^2+1}}{2}}\frac{\Gamma(\sqrt{4L^2+1})}{\Gamma\big(\frac{1}{2}+\frac{\sqrt{4L^2+1}}{2}-iqQ\big)} \big( 1+ \mathcal{O}(\omega r_*)\big).\label{W}
\end{align}
\end{subequations}
Identifying $r$ and $r_*$ at leading order in \eqref{sol ii overlap ii-iii} and \eqref{TK expansions} yields the equations
\begin{subequations}
\begin{align}
    b_1\overset{!}{=}(2i\omega&)^{\frac{1}{2}-\frac{\sqrt{4L^2+1}}{2}}\frac{\Gamma(\sqrt{4L^2+1})}{\Gamma(\frac{1}{2}+\frac{\sqrt{4L^2+1}}{2}-iqQ)}c_2\\
   & b_2\overset{!}{=}(2i\omega)^{\frac{1}{2}+\frac{\sqrt{4L^2+1}}{2}}c_1,\label{c_i fixing}
\end{align}
\end{subequations}
which fix the leading behaviour of the constants $c_1$ and $c_2$ in $\Delta$, $qQ$ and therefore of the solution $\psi_{\omega \ell}^\text{C}$\footnote{\label{c12}Incorporating the behaviour for the $b_i$'s, we find 
    $c_2/c_1=\mathcal{O}\big((\omega M )^{\sqrt{4L^2+1}} \Delta,(\omega M)^{\sqrt{4L^2+1}}(\omega-\omega_\text{I})M \big)$,
i.e. there is an increasing gap for increasing $L^2$.}. Finally, we investigate the behavior of this solution in the outermost portion of region C. For this, we expand the Whittaker functions around $\vert \omega r_*\vert \gg  1$ and obtain\footnote{See for example $13.5$ in \cite{abramowitz1948handbook}.} 
\begin{equation}
\begin{aligned}\label{solution iii large r}
    &\psi_{\omega\ell>0}^\text{C}(\vert \omega r_*\vert \gg 1)= e^{i\omega r_*}\Big\{ c_1 (2i\omega r_*)^{-iqQ}\frac{\Gamma(1+\sqrt{4L^2+1})}{\Gamma\Big( \frac{1}{2}+\frac{\sqrt{4L^2+1}}{2}-iqQ\Big)}\times \Big(1+\mathcal{O}\big( \vert \omega r_* \vert ^{-1}\big) \Big)\Big\}\\
    &+e^{-i\omega r_*}\Big\{c_1 (-2i\omega r_*)^{iqQ}\frac{(-1)^{\frac{1}{2}+\frac{\sqrt{4L^2+1}}{2}}\Gamma(1+\sqrt{4L^2+1})}{\Gamma\Big(\frac{1}{2}+\frac{\sqrt{4L^2+1}}{2}+iqQ \Big)}+c_2(2i\omega r_*)^{iqQ} \Big\}\times \Big(1+\mathcal{O}\big( \vert \omega r_* \vert ^{-1}\big) \Big)
   .
\end{aligned}
\end{equation}
\begin{center}
    \textbf{Region D: asymptotically flat region}
\end{center}
This region extends to radial infinity and captures the right asymptotic behaviour of the radial modes. We may define it via \begin{equation}
    \frac{r}{M}\gtrsim e^{(qQ\omega M)^{-1}}\hspace{1cm}\text{(region D)},
\end{equation}
so that we can safely neglect the terms of order $r^{-2}$ in the radial equation. This region still overlaps with region C. The equation then reads
\begin{equation}
     \frac{d^2\psi_{\omega \ell}^{\text{D}}}{dr_*^2}=\Big\{-\omega^2+\frac{2\omega qQ}{r_*}+\mathcal{O}\Big( \frac{\omega M qQ\ln(r/M)}{r^2}\Big)\Big\}\psi_{\omega \ell>0}^{\text{D}}.
\end{equation}
The solution is a linear combination of the Kummer confluent hypergeometric function and the \textit{confluent hypergeometric function} \cite{Mathematica}, which has approximate behaviour for $\vert \omega r_*\vert \to \infty$
\begin{equation}
    \psi_{\omega \ell}^{\text{D}}=C_{\omega\ell}\  e^{-i\omega r_*}(r_*/\xi)^{iqQ}+ D_{\omega\ell} \ e^{i\omega r_*}(r_*/\xi)^{-iqQ}.
\end{equation}
These constants can be fixed by matching the solution to the large-$r$ form of $\psi^\text{C}_{\omega \ell}$ in \eqref{solution iii large r}. Extracting the coefficients $C_{\omega\ell}$, $D_{\omega\ell}$ in \eqref{BC practice}, and reinserting $L^2=\ell(\ell+1)$ we read off the leading behaviour of the scattering coefficients
\begin{subequations}\label{sc exterior}
\begin{align}
    T_{\omega\ell}^\text{in,I}&=(-1)^{\ell+1}\frac{2^{1+2\ell}(-2i\omega \xi)^{-iqQ}\Delta^\ell(i\omega M)^{\ell+1}\ell!}{(2\ell)!(2\ell+1)!}\ \frac{\Gamma(1+\ell+iqQ)\Gamma\big(\ell+1-i\frac{(\omega-\omega_\text{I})}{\kappa_+}\big)}{\Gamma\big(1-i\frac{\omega-\omega_\text{I}}{\kappa_+}\big)}\Big(1+\mathcal{O}\big(\omega M, \Delta, qQ\big) \Big)\label{T in-I}\\
   R_{\omega\ell}^\text{in,I}&=(-1)^{\ell+1}(-2\omega\xi)^{-2iqQ}\frac{\Gamma(1+\ell+iqQ)}{\Gamma(1+\ell-iqQ)}+\mathcal{O}\big((\omega M )^{2\ell+1}\Delta,(\omega M )^{2\ell+1}(\omega-\omega_\text{I})M)\Big)\label{R in-I},
\end{align}
\end{subequations}
see Fig. \ref{fig:sccoi} for comparison with the numerically obtained results. Some comments are in order: from \eqref{solution iii large r} one sees that $C_{\omega \ell}$ is given by a linear combination of $c_1$ and $c_2$ with coefficients of order $\mathcal{O}(1)$. Since $c_2\ll c_1$, the leading behaviour of $C_{\omega \ell}$ (and therefore of $1/C_{\omega \ell}$ and the scattering coefficient $T_{\omega\ell}^\text{in,I}$) is determined exclusively by $c_1$. The reflection coefficient, on the other hand has leading behaviour $\mathcal{O}(1)$ and subleading corrections depending proportionally on the ratio $c_2/c_1$, which is strongly suppressed for large $\ell$, see footnote \eqref{c12}. Therefore, for increasing $\ell$, corrections to the reflection coefficient \eqref{R in-I} become smaller. We compute such first correction for the $\ell=0$ reflection coefficient explicitly\footnote{In \eqref{R l=0 correction}, the second-order corrections can be any combination which is of third power in the small variables $\omega M$, $\Delta$ and $qQ$. In order to write them compactly we used that $\omega M\lesssim \kappa_+ M \sim \Delta.$ } and find 
\begin{subequations}\label{sc exterior l0}
    \begin{align}
        T_{\omega\ell=0}^\text{in,I} & =-2i\omega r_+(-2i\omega \xi)^{-iqQ}\Gamma(1+iqQ)\Big(1+\mathcal{O}\big(\omega M, \Delta, qQ\big) \Big), \\
        R_{\omega\ell=0}^\text{in,I}&=-(-2\omega \xi)^{-2iqQ}\frac{\Gamma(1+iqQ)}{\Gamma(1-iqQ)}\Big(1-2M^2\omega(\omega-\omega_\text{I})+\mathcal{O}\big(\Delta^3,\Delta^2qQ,\Delta(qQ)^2,(qQ)^3\big) \Big)\label{R l=0 correction},
    \end{align}
\end{subequations}
which can be checked to satisfy their corresponding Wronskian relation \eqref{wronskian inI}.

\section{Semiclassical fluxes}\label{section: semiclassical fluxes}
The analytical expressions for the I/II-scattering coefficients from sec. \ref{section: scattering coeffs} can now be incorporated into the renormalized mode sum expressions for the relevant observables at the event  \eqref{observables EH} and inner horizon \eqref{observables IH}.

\subsection{At the EH}
\begin{figure}
    \centering
    \begin{subfigure}{0.45\textwidth}
        \centering
        \includegraphics[width=\linewidth]{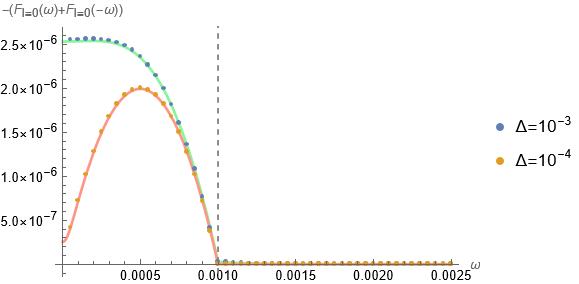}
    \end{subfigure}
    \begin{subfigure}{0.45\textwidth}
        \centering
        \includegraphics[width=\linewidth]{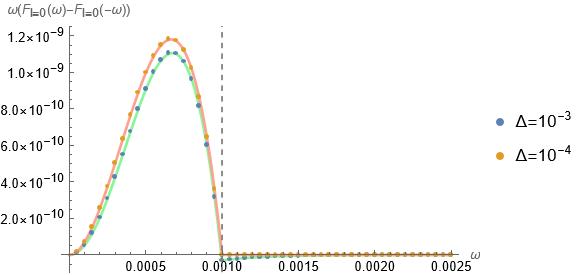}
       
    \end{subfigure}
    \begin{subfigure}{0.45\textwidth}
        \centering
        \includegraphics[width=\linewidth]{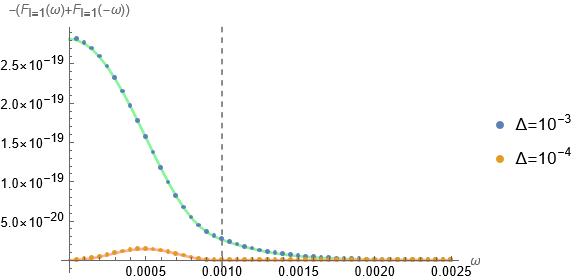}
        
    \end{subfigure}
     \begin{subfigure}{0.45\textwidth}
        \centering
        \includegraphics[width=\linewidth]{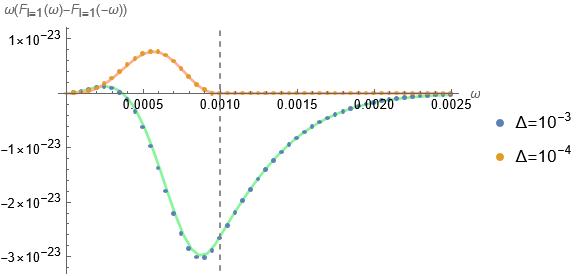}
       
    \end{subfigure}
     \caption{Analytical (solid) and numerical (points) $\ell=0,1$ mode-contributions to  $\langle j_{v}\rangle_\text{U}^{\mathcal{H}^\mathcal{R}}$ (left) and $\langle T_{vv}\rangle_\text{U}^{\mathcal{H}^\mathcal{R}}$ (right) for $qQ=10^{-3}$ and two different extremality parameters. The dashed line indicates the superradiant frequency $\omega_\text{I}$.}\label{fluxes EH}
\end{figure}

The integrand \eqref{Fls} is proportional to $\vert T_{\omega\ell}^\text{up,I}\vert^2$. It scales with $\Delta^{2\ell}$, which suppresses large $\ell$ contributions\footnote{It additionally scales with $((\omega+\omega_\text{I})M)^{2\ell}$, which also suppresses the small-$\omega$ contributions, which in this setup are responsible for the leading behaviour of the integral.}. Inserting the $\ell=0$ leading analytical expressions for the scattering coefficients one finds the integrand in Fig. \ref{fluxes EH}. The integrand localizes more in the superradiant regime as we approach extremality, in which case it peaks at $\omega\approx\omega_\text{I}/2$. The $\ell>0$ mode contributions only enter in the correction terms\footnote{For $\ell>0$ fixed, the integral splits into $\int_0^{\omega_\text{I}}+\int_{\omega_\text{I}}^\infty$. Both integrands are $\mathcal{O}\big(\frac{\omega\mp\omega_\text{I}}{\omega} \vert T_{\pm\omega \ell>0}^{\text{up,I}}\vert^2\big)$ and the effective frequency windows of the integrals are $\mathcal{O}(\omega_\text{I})$ and $\mathcal{O}(\kappa_+-\omega_\text{I})$, respectively.  }, which have been estimated using \eqref{T in-I} and numerically checked, see Fig. \ref{fluxes EH}. The leading contribution for both the current and the renormalized SET comes from the $\ell=0$ mode. Performing the integrations over $\omega$ in \eqref{observables EH} for this mode, we find
\begin{subequations}
\begin{align}
    \langle j_v\rangle_\text{U}^{\mathcal{H}^\mathcal{R}} & =\frac{\pm 1}{16\pi^2 r_+^2}\Big(\frac{4}{3}q^4M^2+\frac{2}{3}q^2\Delta^2 +\mathcal{O}\big(q^6M^4,q^4\Delta,q^2\Delta^3 \big)\Big), \\
     \langle T_{vv}\rangle_\text{U}^{\mathcal{H}^\mathcal{R}} & =\frac{1}{32\pi^2 r_+^2}\Big( \frac{2q^4 M^2}{3}-\frac{\Delta^4}{15 M^2}+\mathcal{O}\big(q^6M^4, q^4\Delta,\Delta^5\big)\Big).
\end{align}
\end{subequations}
The $\pm$ sign corresponds to $Q\gtrless 0$. Both fluxes only involve even powers of the field charge $q$ and decrease to a fixed non-zero value (for $q \neq 0$) as approaching extremality. This can be interpreted as Hawking radiation vanishing in the extremal limit, while superradiant effects remain present and dominate the evaporation mechanism of the black hole. These findings are conceptually in agreement with those of \cite{Klein:2021les}, where a quadratic dependence, $\langle j_v\rangle_\text{U}^{\mathcal{H}^\mathcal{R}} \sim q^2$, was numerically observed in the small charge regime\footnote{Some technical differences exist: their field is (conformally) massive, and the background is given by the Reissner-Nordström-de Sitter (RNdS) geometry. Their analysis focused on larger $\Delta$ regimes for which the quartic behavior is subleading (see, for example, Fig. 2 in that reference) and the quadratic term dominates.}. By the conservation laws \eqref{general forms} and the Hadamard property of the Unruh state we are able to fix the conserved quantities
\begin{subequations}
\begin{align}\label{constant K}
    \mathcal{K} &=\frac{\pm q^4M^2}{12\pi^2}+\frac{\pm q^2\Delta^2}{24\pi^2}+\mathcal{O}\big(q^6M^2, q^4\Delta, \Delta^3 q^2 M^{-2} \big), \\
    \mathcal{L} & =\frac{q^4M^2}{16\pi^2}+\frac{q^2\Delta^2}{24\pi^2}+\frac{\Delta^4}{480\pi^2M^2}+\mathcal{O}\big(q^6M^2, q^4\Delta, \Delta^3 q^2 M^{-2} \big) .
\end{align}
\end{subequations}
to leading order in the small parameters $\Delta$, $\vert qQ\vert$. Since $\mathcal{L}$ has the physical interpretation of energy flux at infinity, its positivity agrees with the expectation that the Unruh state describes an evaporating BH. As $\mathcal{K}$ and $\mathcal{L}$ can be interpreted as the rate of charge loss and mass loss as seen by an asymptotic observer, we see that in the extremal limit $\Delta \to 0$ these rates are related by factor $4/3$ (in the limit of small field charge). Hence, quantum effects drive the BH away from extremality. 

\subsection{At the IH}
\begin{figure}
    \centering
    \begin{subfigure}{0.45\textwidth}
        \centering
        \includegraphics[width=\linewidth]{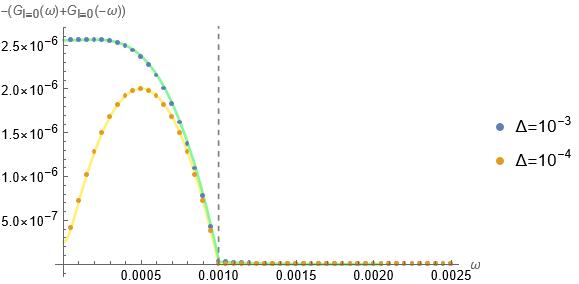}
    \end{subfigure}
    \begin{subfigure}{0.45\textwidth}
        \centering
        \includegraphics[width=1.1\linewidth]{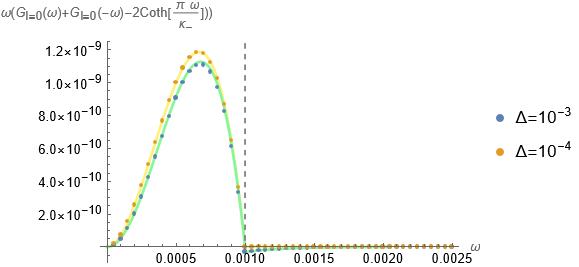}
       
    \end{subfigure}
    \begin{subfigure}{0.45\textwidth}
        \centering
        \includegraphics[width=\linewidth]{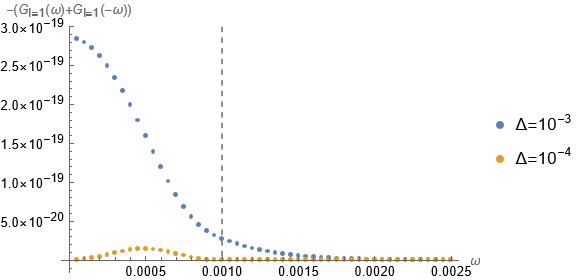}
        
    \end{subfigure}
     \begin{subfigure}{0.45\textwidth}
        \centering
        \includegraphics[width=1.1\linewidth]{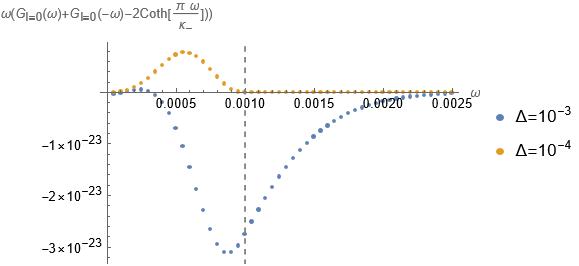}
       
    \end{subfigure}
     \caption{Analytical (solid) and numerical (points) $\ell=0,1$ mode-contributions to  $\langle j_{v}\rangle_\text{U}^{\mathcal{CH}^\mathcal{R}}$ (left) and $\langle T_{vv}\rangle_\text{U}^{\mathcal{CH}^\mathcal{R}}$ (right) for $qQ=10^{-3}$ and two different extremality parameters. The dashed line indicates the superradiant frequency $\omega_\text{I}$.}\label{j IH}
\end{figure}

We evaluate the fluxes at the Cauchy horizon, starting with $\langle j_v \rangle_\text{U}^{\mathcal{CH}^\mathcal{R}}$. We first apply the relations provided in eq. \eqref{relations in up}, and then substitute the analytical leading-order behavior of the $\ell = 0$ scattering coefficients from eqs.~\eqref{sc coeff II} and \eqref{sc exterior l0} into eq. \eqref{current CH}. We expand the integrand in a series for small field charge $qQ$, noting that in this regime $\alpha = 1 + 2\ell - 2iqQ + \mathcal{O}(q^2Q^2)$, where $\alpha$ was defined in \eqref{alpha}. By systematically collecting the leading contributions in the small parameters $\Delta$, $qQ$, and $\omega$, we observe that the term-by-term divergences at $\omega=0$ and $\omega=\omega_\text{II}$ cancel out. The integrand behaves as $\mathcal{O}(\omega^0)$ at small frequencies and is exponentially suppressed at large $\omega$. This analysis yields the solid curve shown in Fig. \ref{j IH}.

The contribution from the $\ell>0$ modes is estimated as follows: first, note that \eqref{G1} is proportional to $\vert T_{\omega^+ \ell}^{\text{up,I}}\vert^2=\mathcal{O}\big( \Delta^{4\ell},\Delta^{2\ell+2}(\omega_\text{I}M)^{2\ell-2}\big)$, in particular it will yield a contribution of subleading order to the $\ell=0$ mode-contribution. For this reason, we can already neglect it at this level. We substitute the analytical scattering coefficients into the remaining terms and, as with the $\ell=0$ modes, expand the sums $G_{\ell>0}^i(\omega) + G_{\ell>0}^i(-\omega)$ for $i=2,3$ in a series for small field charge $qQ$. Each term diverges quadratically as $\omega \to 0$, and this behavior is independent of $\ell$. When the terms are combined, the divergences cancel out, and the leading-order contributions vanish entirely. Therefore, we have explicitly found that the $\ell=0$ mode is responsible for the leading behaviour, as also verified by the numerical results shown in Fig.~\ref{j IH}.\footnote{Note that no analytical results are shown in Fig.~\ref{j IH} for the contribution of the $\ell = 1$ mode. The reason is that the integrand at the IH is sensitive to the distinction between $\kappa_+$ and $\kappa_-$, see \eqref{observables IH}, which is beyond the regime of our approximations. The strong cancellations in the integrand (at $\omega=0$ and $\omega=-\omega_\text{II}$) are also sensitive to the subleading corrections in $\Delta$, so our analytical expressions (only to leading order in $\Delta$) can not capture the correct behaviour.} Performing the integration over $\omega$ for the $\ell = 0$ mode, we thus obtain
\begin{equation}\label{current IH final}
    \langle j_v \rangle_\text{U}^{\mathcal{CH}^\mathcal{R}}=\frac{\pm 1}{16\pi^2 r_-^2}\Big(\frac{4}{3}q^4M^2+\frac{2}{3}q^2\Delta^2 +\mathcal{O}\big(q^6M^4, \Delta^2 q^4M^2, \Delta^3 q^2 \big)\Big).
\end{equation}
The $\pm$ sign corresponds to $Q\gtrless 0$. We notice that, within our approximation, $\langle j_v \rangle_\text{U}^{\mathcal{CH}^\mathcal{R}} = \frac{\mathcal{K}}{r_-^2}$, with $\mathcal{K}$ as determined in \eqref{constant K}. With \eqref{general forms}, one concludes that the $u$ component of the charge flux is subleading to the $v$ component, i.e., of the order of the correction terms in \eqref{current IH final}.

An analogous calculation for the renormalized SET shows that the mode-sum is again dominated by the $\ell=0$ modes and yields
\begin{equation} \label{RSET IH final}
    \langle T_{vv}\rangle_\text{U}^{\mathcal{CH}^\mathcal{R}}=\frac{1}{32\pi^2 r_-^2}\Big( \frac{2q^4 M^2}{3}-\frac{\Delta^4}{15 M^2}\Big)+\mathcal{O}\big( q^4 \Delta, \Delta^5 M^{-4}, \Delta^4 q^2 M^{-2}, q^6M^2\big)
\end{equation}
which agrees with the available results for the real scalar field \cite{Zilberman:2021vgz} in the limit $ qQ \to 0$. We see that both positive and negative sign can occur, but predict positivity for $q\neq 0$ sufficiently close to extremality. Again, we find that, within our approximation, $\langle T_{vv}\rangle_\text{U}^{\mathcal{CH}^\mathcal{R}} = - \frac{\mathcal{L}}{r_-^2} + \frac{\mathcal{K} Q}{r_-^3}$, from which we can conclude, with \eqref{general forms}, that the $uu$ component of the stress tensor is subleading to the $vv$ in our approximation. The same phenomenon was already observed for the real scalar field \cite{Zilberman:2021vgz}.\footnote{
Regarding the comparison with Kerr BHs, two relevant points should be noted: (1) near extremality, it was found \cite{Zilberman:2022aum} that the polar fluxes satisfy the same condition $\langle T_{uu}\rangle^{\mathcal{CH}^\mathcal{R}}\ll \langle T_{vv}\rangle^{\mathcal{CH}^\mathcal{R}}$, with both fluxes vanishing in the extremal limit, and (2) in the extremal limit, off-pole (where superradiant effects are present), a weaker statement was predicted \cite{Zilberman:2022iij}. Specifically, $\langle T_{uu}\rangle_\text{U}^{\mathcal{CH}^\mathcal{R}} < \langle T_{vv}\rangle_\text{U}^{\mathcal{CH}^\mathcal{R}}$ in  a coordinate system that maintains a regular azimuthal coordinate across the inner horizon.} In this sense, in the near-extremal approximation, only negligible scattering takes place in the BH interior, i.e., the fluxes at the IH and at the EH coincide at leading order. Intuitively, this result can be interpreted as the horizons being very close to each other in the near-extremal regime, leading to negligible scattering between them.

Considering the above fluxes in a coordinate system which can be extended regularly across the Cauchy horizon, i.e., using the tensor transformation law to obtain $\langle j_V \rangle_\text{U}^{\mathcal{CH}^\mathcal{R}}$ and $\langle T_{VV}\rangle_\text{U}^{\mathcal{CH}^\mathcal{R}}$, we see that the former diverges as $V^{-1}$ and the latter as $V^{-2}$. This divergent behavior indicates an instability of the Cauchy horizon and provides a strong motivation for the study of backreaction effects, for which we perform first steps in the next section.

\section{Physical consequences}\label{section: discussion}

Finally, we analyze the effects of local backreation of the charged quantum field onto the spacetime. This proceeds by inserting the explicit expectation values from sec. \ref{section: semiclassical fluxes} into the right hand side of the semi-classical Einstein-Maxwell equation \eqref{semicl EN}. On the left hand side we allow for a spherically symmetric metric and field strength tensor
\begin{equation}\label{spherical symmetry ansatz}
    g=-e^{\sigma(u,v)}du dv+r^2(u,v)d\Omega^2, \hspace{1.5cm}F=-\frac{Q(u,v)}{2 r^2(u,v)}e^{\sigma(u,v)}du\land dv,
\end{equation}
where $r, \sigma$ and $Q$ are apriori unknown functions of the double null coordinates $u$ and $v$. For this choice, one can show that $T_{vv}^\text{EM}=0$ and that $\nabla^\mu F_{\mu v}=\partial_v Q(u,v)r^{-2}$. Then, the $vv$-component of the semiclassical Einstein-Maxwell equations \eqref{semicl EN} becomes
\begin{subequations}\label{vv component EE}
\begin{align}
  G_{vv} & =8\pi\langle T_{vv}\rangle_\Psi, \label{5.7a}\\
  \partial_v Q(u,v) & =-4\pi r^2 \langle j_v\rangle_\Psi. \label{5.7b}
\end{align}
\end{subequations}
Moreover, since $g_{vv}=0$, the addition of a cosmological constant does not modify any of the above equations. In particular, our analysis closely follows the one performed in \cite{Klein:2021ctt}, where backreaction effects are studied in the vicinity of the Cauchy horizon in a dynamical spherically symmetric spacetime with non-vanishing cosmological constant. 
\begin{figure}
    \centering
    \begin{subfigure}{0.47\textwidth}
        \centering
        \includegraphics[width=.8\linewidth]{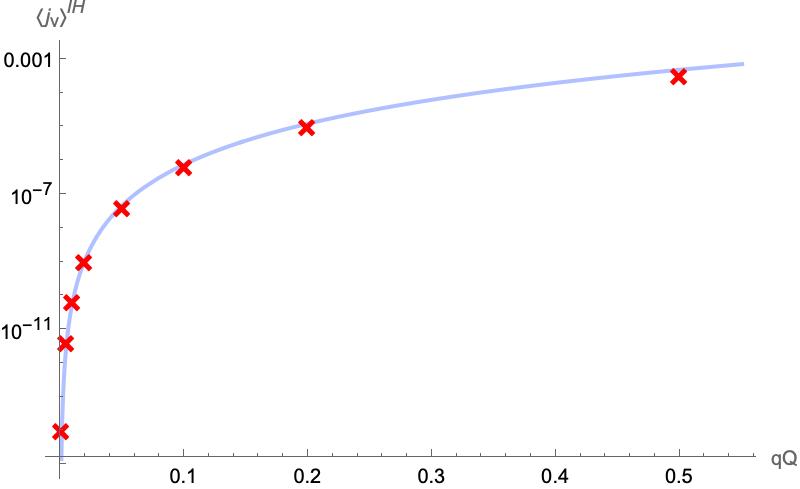}
    \end{subfigure}
    \begin{subfigure}{0.47\textwidth}
        \centering
        \includegraphics[width=.8\linewidth]{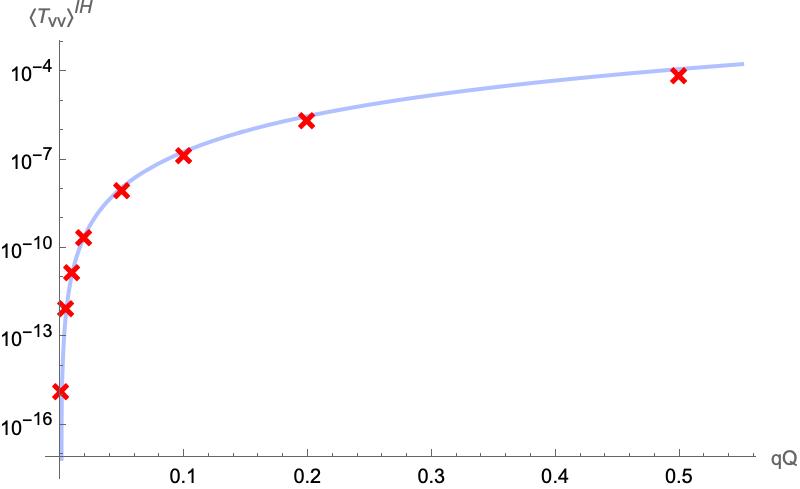}
       
    \end{subfigure}
     \caption{Numerical results for $\langle j_v\rangle_\text{U}^\mathcal{CH^\mathcal{R
    }}$ (left) and $\langle T_{vv}\rangle_\text{U}^\mathcal{CH^\mathcal{R
    }}$ (right) for fixed near-extremality parameter $\Delta=10^{-3}$ as a function of the field charge, ranging from $qQ=10^{-3}$ to $qQ=0.5$ (and $Q\approx 1$ fixed). The solid line corresponds to the analytical approximations \eqref{current IH final} and \eqref{RSET IH final}, which were derived for $qQ\ll1$. While the analytical approximation  is very good in that regime ($<1\%$ for $qQ=10^{-3}$), the deviations for larger values of $q$ are perceptible ($29\%$ for $qQ=0.5$). Nevertheless, the qualitative features (positive, monotonously increasing) are captured by the approximation. For all field charges (not necessarily small), the fluxes were dominated by the $\ell=0$ mode, suggesting that this is a feature inherent to the near-extremal regime (and independent of the field charge). } \label{largeQq}
\end{figure}

In the following, we study the weak-backreaction effects of the RN stress-tensor and current of a charged scalar field in the Unruh state, in the vicinity of the IH \cite{Zilberman:2019buh,Klein:2021ctt}.  We decompose the unknown functions $r$ and $\sigma$ into  their background RN part $r_0$, $\sigma_0$ and a perturbation $\delta r$, $\delta \sigma$. In the weak-backreaction regime we shall assume that $\langle j_v\rangle_\text{U}$ and $\langle T_{vv}\rangle_\text{U}$ are still well-approximated by their unbackreacted analogs. The perturbations are assumed to be weak enough so that in the vicinity of the IH we can substitute $r^2$ by its background value $r_-^2$ in \eqref{5.7b}, obtaining

\begin{equation}\label{weak BR}
    \partial_v Q(u,v)=-4\pi r_- ^2 \langle j_v\rangle_\text{U} ^{\mathcal{CH}}=-4\pi \mathcal{K}.
\end{equation}
The constant $\mathcal{K}$ was defined in \eqref{general forms} and found to have the same sign as $Q$, see \eqref{constant K}. While the analytical expression given there is only valid in the regime of small field charge, its qualitative features (in particular, its sign) are expected (and numerically confirmed) to be valid for generic field charges, see Fig. \ref{largeQq}.  This can be explained as follows: in the near-extremal regime, the EH fluxes are determined entirely by superradiant effects. For small field charges $\vert qQ\vert \ll 1$ we showed (see \eqref{current EH}, \eqref{current IH final}) that EH and IH fluxes coincide to leading order. For larger charges, the superradiant frequency band ($\omega_\text{I}\approx q$) widens, allowing more modes to contribute. However, these additional modes do not scatter significantly in the interior, as the reflection coefficient is exponentially suppressed for $\omega\gtrsim \kappa_+$ (see \eqref{sc coeff II} and discussion below, so that everything "goes through"\footnote{In terms of the integrand functions $G_\ell(\omega)$ and $F_\ell(\omega)$ entering the IH \eqref{observables IH}, resp. EH \eqref{observables EH} mode-sums, one can show that the combination $G_\ell(\omega)+G_\ell(-\omega)$ reduces, at leading order, to its EH counterpart $F_\ell(\omega)+F_\ell(-\omega)$, and similarly for the RSET.}. As a result, the EH fluxes propagate to the IH without leading-order variations, so that the IH fluxes remain governed by superradiance, regardless of the field charge magnitude. In particular, this means that in the near-extremal regime (and for generic field charges), backreaction effects tend to discharge the black hole interior ($r<r_-$) , agreeing with \cite{Klein:2021ctt}. On the other hand, the semiclassical Einstein equation \eqref{5.7a} may be rewritten in terms of the unkown functions $r$ and $\sigma$ as
\begin{equation}
   \partial_v^2 r(u,v)-\partial_v\sigma(u,v) \partial_v r(u,v)=-4\pi r(u,v) \langle T_{vv}\rangle_\Psi.
\end{equation}
Restricting to the IH vicinity and substituting $r(u,v)\to r_-$ in the final expression leads to \begin{equation}
    \partial_v^2 \delta r (u,v)+\kappa_- \partial_v \delta r+\mathcal{O}\big( \partial_v \delta \sigma 
    \partial_v  \delta r\big)=- 4\pi r_-\langle T_{vv}\rangle_\text{U}^{\mathcal{CH}^\mathcal{R}}.
\end{equation}
Neglecting the terms quadratic in the perturbation, the solution to this ordinary differential equation is a linear combination of the homogeneous and inhomogeneous solution
\begin{equation}
         \partial_v \delta r (u,v)=C\cdot e^{-\kappa_- v}-4\pi \frac{r_-}{\kappa_-}\langle T_{vv}\rangle_\text{U}^\mathcal{CH}.
\end{equation}
We argue that in the near-IH regime ($v\to \infty)$, the exponential term can be dropped. Assuming that this is still the case for small (but fixed) $\kappa_-$, one obtains the relation \cite{Klein:2021ctt,Zilberman:2019buh}
\begin{equation}\label{v derivative r}
    \partial_v \delta r(u,v) =-4\pi \frac{r_-}{\kappa_-}\langle T_{vv}\rangle_\text{U}^\mathcal{CH}.
\end{equation}
A negative (positive) sign in the right hand side of the above equation is related to spontaneous contraction (expansion) of the IH \cite{Zilberman:2019buh}. Recalling that our results for $\langle T_{vv}\rangle_\text{U}^\mathcal{CH}$ approach a (positive) constant value in the extremal-limit ($\kappa_-\to 0$), the appearance of $\kappa_-$ in the denominator in \eqref{v derivative r} indicates that the drift in $r(u,v)$ from its background accumulates faster in scenarios which are originally closer to extremality. 

To account for backreaction effects on the mass within the black hole interior, a localized region of the spacetime, we first introduce a quasi-local notion of mass or energy. A suitable candidate in spherically symmetric spacetimes is the mass function proposed by Poisson and Israel \cite{Poisson:1990eh}, given by
\begin{equation}
    M_{\text{P-I}}(u,v)\equiv \frac{r(u,v)}{2}\big(1-g^{\alpha\beta}\partial_\alpha r(u,v)\partial_\beta r(u,v)\big)+\frac{Q^2(u,v)}{2r(u,v)}.
\end{equation}
This definition of mass is closely related to the Misner-Sharp\footnote{In the case of the Reissner-Nordström solution, $M_{\text{P-I}}$ reduces to the parameter $M$, while the Misner-Sharp mass reduces to $M - Q^2/r$, where $M$ and $Q$ are the background parameters entering the metric \eqref{RN}.} mass \cite{Misner:1964je}. We find
\begin{equation}
    M_\text{P-I}(u,v)=\frac{r_0+\delta r}{2}\Big\{1-e^{\delta \sigma}f(r_0)+2 e^{-\delta \sigma}\big(\partial_u \delta r-\partial_v \delta r +\mathcal{O}\big(\partial_u \delta r\partial_v \delta r\big)\big) \Big\}+\frac{Q^2(u,v)}{2(r_0+\delta r)}.
\end{equation}
Taking a $v-$derivative of this quantity, evaluating it at the IH (where $f(r_0),\partial_v f(r_0)\to 0$) and collecting the terms to first order in the perturbation yields
\begin{equation}
\begin{aligned}
    \partial_v   M_\text{P-I}(u,v)&=\partial_v \delta r\big(-r_- \kappa_-+\mathcal{O}\big( 
    \partial_y \delta r,\partial_v^2 \delta r,\partial_v \sigma \partial_y \delta r\big) \big) +\frac{Q\partial_v Q(u,v)}{r_-}\\
    &=4\pi r_-^2\Big( \langle T_{vv}\rangle_\text{U}^{\mathcal{CH}^\mathcal{R}}- \frac{Q}{r_-} \langle j_{v}\rangle_\text{U}^{\mathcal{CH}^\mathcal{R}}\Big)
\end{aligned}
\end{equation}
with $y\in\{u,v \}$. Inserting the explicit expectation values \eqref{current IH final}, \eqref{RSET IH final} we find $\partial_v M_{\text{P-I}}(u,v)=-4\pi \mathcal{L}$, where the constant $\mathcal{L}$ was defined in \eqref{general forms}. In particular, since the constant $\mathcal{L}$ was found to be strictly positive, the BH interior is strictly evaporating. Furthermore, in the extremal limit we again find the same ratio $4/3$ between the decay rates $\frac{\del_v Q}{Q}$ and $\frac{\partial_v M_{\text{P-I}}}{M_{\text{P-I}}}$ of the local charge and quasi-local mass that we already found for the asymptotic values of charge and mass in the exterior. The direct relationship between the quasi-local quantities $\partial_v Q(u,v)$, $\partial_v M_\text{P-I}(u,v)$ at the Cauchy horizon and the fluxes at infinity, $\mathcal{K}$ and $\mathcal{L}$, is a peculiarity of the near-extremal regime and we do not expect it to hold generically.

\section{Concluding remarks}
In this paper we studied the semiclassical fluxes associated with a charged scalar field in the Unruh state in a near-extremal charged BH spacetime, described by the Reissner-Nordström solution \eqref{RN} to the Einstein-Maxwell equations. 

The analysis is built upon recent work on charged scalar fields in RN(dS) backgrounds \cite{Klein:2021ctt,Klein:2021les,Balakumar:2022yvx} where explicit forms for the renormalized expectation values of relevant observables are presented as mode sums which can be evaluated at the horizons. 

The key ingredient to the mode-sums are the scattering coefficients of the Boulware modes. Obtaining these analytically (by generalizing the results of \cite{Zilberman:2021vgz} for the real scalar field) was one of the central aspects of this work: in the interior region, they can be solved quite directly to leading order in the small (extremality) parameter $\Delta$. Solving the scattering problem in the exterior region I is less trivial and involves subpartitioning the region into subregions where certain approximations can be made. In particular, we made the supplementary assumption of small field charge $q Q \ll 1$. In both regions we verified our analytical approximations by comparison with numerical results.

Using the analytical approximations for the scattering coefficients, we obtained analytical approximations for $\langle j_v\rangle_\text{U}$ and $\langle T_{vv}\rangle_\text{U}$ at the event and Cauchy horizon. From the fact that these coincided at the two horizons (within our approximation) and the conservation of the current and the stress tensor, we could conclude that $\langle j_u\rangle_\text{U}$ and $\langle T_{uu}\rangle_\text{U}$ at the Cauchy horizon are suppressed with respect to $\langle j_v\rangle_\text{U}$ and $\langle T_{vv}\rangle_\text{U}$ by further factors of $\Delta$ and/or $q Q$. We find that both positive and negative signs for the energy flux at both horizons may occur but that sufficiently close to extremality the positive sign dominates. This is in contrast to the real scalar field, where the semiclassical energy fluxes are negative in the near extremal regime and vanish in the extremal limit, i.e., $\langle T_{vv}\rangle^{\mathcal{H}^\mathcal{R}/\mathcal{CH}^\mathcal{R}}\to 0^-$ \cite{Zilberman:2021vgz}. 

We discussed the physical consequences of incorporating quantum effects (i.e. the above expectation values) into the right hand side of the semiclassical Einstein-Maxwell equations \eqref{semicl EN} in the context of weak backreaction. We found that in the near-extremal regime, quantum effects discharge the black hole interior and drive it away from extremality, in agreement with \cite{Klein:2021ctt} and resembling the results in \cite{Klein:2024sdd} (where quantum effects are found to decrease the angular momentum of rotating BHs).

With respect to the previous study \cite{Zilberman:2021vgz} of the real scalar field (which our results generalize), the main novelty is the consideration of superradiance, which is physically the generic situation. However, superradiance complicates the analysis considerably, and in order to nevertheless arrive at analytical expressions, we made the assumption of small field charge, $q Q \ll 1$, to limit the extent of the superradiant frequency range. It would be desirable to get rid of this assumption, in particular in view of extending the analysis to the physically more relevant case of near-extremal rotating BHs. There, the quantity analogous to $q Q$ is $m/2$ (with $m$ the azimuthal quantum number of the modes), which can obviously not be assumed to be a small parameter.

\section*{Acknowledgments}
M.A. thanks Christiane Klein, Philipp Dorau, Noa Zilberman and Amos Ori for valuable discussions that contributed positively to the development of this paper. We also thank Marc Casals for pointing out the connection to previous work \cite{Page:1976df} in the matching of asymptotic expansions. We are grateful to Stefan Hollands for his hospitality during M.A.'s stay in Leipzig and 
for discussions that inspired the initial idea for this project. This work has been funded by the Deutsche Forschungsgemeinschaft (DFG) under Grant No. 406116891 within the Research Training Group RTG 2522/1, with additional support for M.A. provided by the Max Planck Society through the Max Planck-Israel Program.

\appendix
\section{Scattering in the interior region}\label{AppendixA}
The general solution to \eqref{general II radial equation} is a linear combination of two solutions
    $h_{\omega \ell}^{(+)\text{in,II}}=c_1 \psi^1+c_2\psi^2$ (dropping the indices for convenience), given by

\begin{equation}
\begin{aligned}
     \psi^1&=-i\sech^{-1}(\Tilde{r}_*)(-1+\tanh(\Tilde{r}_*))^{\frac{1}{2}+\frac{i}{2}(\Tilde{\omega}-2qQ)}(1+\tanh(\Tilde{r}_*))^{\frac{1}{2}-\frac{i\Tilde{\omega}}{2}}\\
     &\times{}_2F_1\Big[\frac{1}{2}(1+\alpha),\frac{1}{2}(1-\alpha);-2iqQ+\alpha+i(\Tilde{\omega}-2qQ);\frac{1}{1+e^{2\Tilde{r}_*}} \Big]
\end{aligned}
\end{equation}
\begin{equation}
    \begin{aligned}
        \psi^2&=-i\sech^{-1}(\Tilde{r}_*)(-1+\tanh(\Tilde{r}_*))^{\frac{1}{2}+\frac{i}{2}(\Tilde{\omega}-2qQ)}(1+\tanh(\Tilde{r}_*))^{\frac{1}{2}-\frac{i\Tilde{\omega}}{2}}(1-\tanh(\Tilde{r}_*))^{i(2qQ-\Tilde{\omega})}\\
     &\times(-1/2)^{i(2qQ-\Tilde{\omega})}{}_2F_1\Big[\frac{1}{2}(1+\alpha)+i(2qQ-\Tilde{\omega})),\frac{1}{2}(1-\alpha)-i\Tilde{\omega};1-i(\Tilde{\omega}-2qQ);\frac{1}{1+e^{2\Tilde{r}_*}} \Big].
    \end{aligned}
\end{equation}
${}_2 F_1(a,b;c;z)$ denotes the hypergeometric function. The constant $\alpha$ is defined as 
\begin{equation}\label{alpha}       
  \alpha\equiv\frac{+iqQ(1+2\ell)}{-2 q^2 Q^2 + \sqrt{q^2 Q^2 (-1 - 4 \ell(\ell+1) + 4 q^2 Q^2)}}.
\end{equation}

In order to extract the scattering coefficients from the general solution one imposes the boundary conditions \eqref{(+)inII modes} at the EH ($r_*\to -\infty$), which allows to fix $c_1$ and $c_2$, and then reads off the scattering coefficients by taking the solution to the IH ($r_*\to+\infty$). 
 In the EH limit $\Tilde{r}_*\to -\infty$, one may substitute
\begin{equation*}
    (-1+\tanh(\Tilde{r}_*))\to -2, \hspace{1cm} (1+\tanh(\Tilde{r}_*))\to2 e^{2\Tilde{r}_*}, \hspace{1cm}\sech^{-1}(\Tilde{r}_*)\to \frac{1}{2}e^{-\Tilde{r}_*}
\end{equation*}
and expand the hypergeometric function in a series. Imposing the defining inII-Boulware conditions \eqref{(+)inII modes} yields two equations which allow to fix the coefficients $c_1$, $c_2$. The scattering coefficients can then be read off by taking $\psi^1$ and $\psi^2$ to the IH. In the limit $\Tilde{r}_*\to +\infty$, the hyperbolic functions have asymptotic behaviour
\begin{equation*}
    (-1+\tanh(\Tilde{r}_*))\to -2e^{-2\Tilde{r}_*}, \hspace{1cm} (1+\tanh(\Tilde{r}_*))\to2 , \hspace{1cm}\sech^{-1}(\Tilde{r}_*)\to \frac{1}{2}e^{\Tilde{r}_*}.
\end{equation*}
The hypergeometric functions are also expanded in a series about $\Tilde{r}_*=+\infty$, in this case one has $$ {}_2F_1\Big[a,b;c;\frac{1}{1+e^{2\Tilde{r}_*}}\Big]=1+\mathcal{O}(e^{-2\Tilde{r}_*})$$ for any $a$, $b$, $c$. Comparison with \eqref{(+)inII modes} yields the scattering coefficients \eqref{sc coeff II}. One may verify that these reduce to the scattering coefficients for the uncharged scalar \cite{Zilberman:2021vgz} in the limit $qQ\to 0$ and that they satisfy the Wronskian relation \eqref{wronskian II exact} to leading order in $\Delta$, namely
\begin{equation}\label{wronskian approximate}
    1=\frac{\omega-\frac{2qQ\Delta}{M}}{\omega}\big(\vert T_{\omega \ell} ^{\text{in,II}}\vert^2-\vert R_{\omega \ell} ^{\text{in,II}}\vert^2 \big).
\end{equation}

\end{document}